\documentclass[12pt]{article}
\usepackage{amsmath}
\usepackage{amssymb}
\usepackage{amsfonts}
\usepackage{graphicx}
\usepackage[utf8]{inputenc}			% inspire bibs
\usepackage{aas_macros}				% ads bibs
\usepackage{slashed}       % \slashed{k}
\usepackage{xspace}			% for spacing in macros

\usepackage[dvipsnames]{xcolor}
\usepackage{cite}          % group cites (conflict: collref)
\usepackage{float}
\usepackage{multirow}

\usepackage{booktabs}      % professional tables
\usepackage{multirow}      % multirow elements in a table
\usepackage{tikzfeynman}   % Flip's rules Feynman Diagrams

\usepackage[margin=2cm]{geometry}   % margins
\graphicspath{{figures/}}			% figure folder
\numberwithin{equation}{section}    % set equation numbering

\newcommand{\acro}[1]{\textsc{\MakeLowercase{#1}}\xspace}

\renewcommand{\tilde}{\widetilde}   % tilde over characters
\renewcommand{\vec}[1]{\mathbf{#1}} % vectors are boldface
\newcommand{\ModelI}{$\mathbb{C}e\mu$}
\newcommand{\ModelII}{$\mathbb{C}\mu\tau$}
\newcommand{\ModelIII}{$\mathbb{R}e\mu$}
\newcommand{\ModelIV}{$\mathbb{R}\mu\tau$}

\newcommand{\beq}{\begin{equation}}
\newcommand{\eeq}{\end{equation}}

\newcommand{\hc}{\mbox{h.c.}}

\newcommand{\lp}{\left(}
\newcommand{\rp}{\right)}

\newcommand{\abs}[1]{\left| #1\right|}
\newcommand{\order}[1]{\mathcal{O}\lp #1\rp}

\let\oldenumerate\enumerate
\renewcommand{\enumerate}{
  \oldenumerate
  \setlength{\itemsep}{1pt}
  \setlength{\parskip}{0pt}
  \setlength{\parsep}{0pt}
}

\let\olditemize\itemize
\renewcommand{\itemize}{
  \olditemize
  \setlength{\itemsep}{1pt}
  \setlength{\parskip}{0pt}
  \setlength{\parsep}{0pt}
}

\newcommand{\email}[1]{\href{mailto:#1}{#1}}

\newenvironment{institutions}[1][2em]{\begin{list}{}{\setlength\leftmargin{#1}\setlength\rightmargin{#1}}\item[]}{\end{list}}

\usepackage{fancyhdr}		% to put preprint number

\usepackage[
	colorlinks=true,
	citecolor=green!50!black,
	linkcolor=NavyBlue!75!black,
	urlcolor=green!50!black,
	hypertexnames=false]{hyperref}

\fancypagestyle{firststyle}{
\rhead{\footnotesize \texttt{UCR-TR-2019-FLIP-ECS-1022}}}

\usepackage{etoc}

\begin{document}

\thispagestyle{firststyle} %% to include preprint

\begin{center}

 {\Large \bf Exotic Lepton-Flavor Violating Higgs Decays}

 \vskip .7cm

  { \bf Jared A.~Evans$^{a}$, Philip Tanedo$^{b}$, Mohammadreza Zakeri$^{c}$}
 \\ \vspace{-.2em}
 { \tt
  \footnotesize
  \email{jaredaevans@gmail.com},
  \email{flip.tanedo@ucr.edu},
  \email{mzake001@ucr.edu}
 }

 \vspace{-.2cm}

 \begin{institutions}[2.25cm]
  \footnotesize
  $^{a}$ {\it
    Department of Physics,
    University of Cincinnati,
    OH 45221
   }
  \\
  \vspace*{0.05cm}
  $^{b}$ {\it
    Department of Physics \& Astronomy,
    University of  California, Riverside,
    {CA} 92521
   }
  \\
  \vspace*{0.05cm}
  $^{c}$ {\it
    Institute of Theoretical Physics,
    Chinese Academy of Sciences, Beijing 100190,
    {China}
   }
 \end{institutions}

\end{center}

\begin{abstract}
 \noindent
Exotic Higgs decays are promising channels to discover new physics in the near future.  We present a simple model with a new light scalar that couples to the Standard Model through a charged lepton-flavor violating interaction.  This can yield exciting new signatures, such as $h \to e^+ e^+ \mu^-\mu^-$, that currently have no dedicated searches at the Large Hadron Collider.  We discuss this model in detail, assess sensitivity from flavor constraints, explore current constraints from existing multi-lepton searches, and construct a new search strategy to optimally target these exotic, lepton-flavor violating Higgs decays.
\end{abstract}

\small
\setcounter{tocdepth}{2}
\tableofcontents
\normalsize
\clearpage

%%%%%%%%%%%%%%%%%%%%%%%%%
\section{Introduction}

The Large Hadron Collider (\acro{LHC}) has produced millions of Higgs bosons since the 2012 discovery by the \acro{ATLAS} and \acro{CMS} collaborations~\cite{ATLAS:2012ae,Chatrchyan:2012tx}. This growing sample allows for searches probing exotic Higgs decays, which may provide the pathway to uncovering new physics~\cite{Curtin:2013fra}. Since the Higgs vacuum expectation value (\acro{VEV}) is the order parameter of electroweak symmetry breaking, the Higgs field has a special role in the possible ways that the Standard Model can couple to otherwise inaccessible new particles. Our current data about the Higgs sector may easily accommodate an $\order{10\%}$ branching fraction into exotic final states.  In fact, many possible exotic final states could be discoverable even if only a few Higgses decay that way over the lifetime of the \acro{LHC}, e.g.~\cite{Curtin:2013fra,Izaguirre:2018atq}.  

Hidden sectors \cite{Battaglieri:2017aum} can provide a very simple origin for an exotic Higgs decay.  One of the most compelling reasons to consider the addition of a hidden sector onto the Standard Model is dark matter.   For example, dark matter may interact with visible matter through a lower-mass mediator~\cite{Pospelov:2007mp,Feng:2008ya, Alexander:2016aln}. The dark matter--mediator coupling can be fixed to yield the observed dark matter abundance through thermal freeze out.  On the other hand, the couplings of the mediator to visible matter only needs to be large enough to account for the observed dark matter relic abundance. This can easily be small enough to avoid laboratory-based experimental bounds \cite{Evans:2017kti,Evans:2019vxr}.  Independent of any connection to dark matter, the possible existence of low-mass particles that interact weakly with visible matter may be tested through their interactions with the Higgs. In fact, for many low-energy coupling structures such a connection to the Higgs may be mandated by electroweak gauge invariance.

The discovery of lepton flavor violating (\acro{LFV}) couplings in the charged lepton sector would be a striking indication of physics beyond the Standard Model.  The possibility of additional \acro{LFV} contributions beyond neutrino mixing is especially tantalizing in the wake of various flavor physics anomalies involving leptons.  Among these anomalies are the lepton flavor universality violation in $R_{K^{(*)}}$ and  $R_{D^{(*)}}$~\cite{Aaij:2014ora,Aaij:2015yra}, the proton radius puzzle~\cite{Pohl:2010zza}, and the long standing anomalous magnetic moment of the muon \cite{Bennett:2006fi}. All of these have received various hidden sector explanations, see e.g.~\cite{Pospelov:2008zw, Barger:2010aj,Datta:2017pfz,Asadi:2018wea,Greljo:2018ogz,Robinson:2018gza}. There are, of course, many other observables that have so far proven consistent with Standard Model expectations, such as muonium oscillation, $\mu^- \to e^- e^+ e^-$~\cite{Bellgardt:1987du}, or $\tau \to \mu \gamma$~\cite{Aubert:2009ag}, that provide stringent constraints on possible new physics.   

A largely unexplored scenario is that of a light, spin-0 particle $\varphi$ that interacts with the Standard Model through \acro{LFV} couplings to charged leptons~\cite{Galon:2016bka}. This has been studied \cite{Galon:2016bka} in the context of fitting the $\gamma$-ray excess in the galactic center~\cite{Goodenough:2009gk,
 Hooper:2010mq,
 Abazajian:2010zy,
 Boyarsky:2010dr,
 Hooper:2011ti,
 Abazajian:2012pn,
 Gordon:2013vta,
 Hooper:2013rwa,
 Huang:2013pda,
 Okada:2013bna,
 Macias:2013vya,
 Abazajian:2014fta,
 Daylan:2014rsa,
 Zhou:2014lva,
 Calore:2014xka,
 Calore:2014nla,
 Calore:2015nua,
 TheFermi-LAT:2015kwa}.\footnote{For the current status of this anomaly, we refer to recent statistical analyses supporting either a possible dark matter~\cite{Leane:2019xiy} or point source~\cite{Chang:2019ars} interpretation.  The proposed collider search in this manuscript is independent of the ultimate interpretation of the $\gamma$-ray excess. }
When this scalar is lighter than the Higgs, it is possible for the Higgs to decay through an exotic four-lepton channel, where $h\to \ell^+\ell'^-\varphi \to 4\ell$ or $h\to \varphi^*\varphi \to 4\ell$.  In the case of prompt $\varphi$ decays, this exciting signature does not yet have any dedicated search. Existing searches for displaced vertices probe down to $\mathcal O(100~\mu\text{m})$, corresponding to coupling strengths $\lesssim 10^{-6}$ for an 10~GeV mediator; however, they are not optimized for this signature.  Further, despite the simplicity of the model, there are exotic four-lepton signatures, e.g.~$e^+e^+\mu^-\mu^-$, that can naturally emerge easily, but that have thus far skirted systematic signature classification programs, see e.g.~\cite{Curtin:2013fra,Craig:2016rqv,Kim:2019rhy}.

In addition to the potential collider signatures, this model of a scalar with \acro{LFV} couplings could explain the outstanding discrepancy in the muon anomalous magnetic moment~\cite{Bennett:2006fi,Galon:2016bka}.  A separate dedicated collider study of this model focused on a mediator that is heavier than dark matter and the decay of the Standard Model-like Higgs boson to $\tau\mu$~\cite{Galon:2017qes}. Recent complementary studies have explored displaced vertices~\cite{Heeck:2017xmg} and signals at future $e^+e^-$ colliders~\cite{Dev:2017ftk}.  We note that our focus is distinct from the extensively studied flavorful dark sector scenarios in which both the mediator and dark matter carry flavor charges, e.g.~\cite{Kile:2011mn,
 Batell:2011tc,
 Kamenik:2011nb,
 Agrawal:2011ze}.
 
In this paper we examine the collider phenomenology of a scalar that is produced in decays of the Higgs and then subsequently decays through an exotic \acro{LFV} coupling.  The low-energy model is detailed in Section~\ref{sec:model}.  Section~\ref{sec:flavor} explores flavor constraints both when only a single off-diagonal coupling is present, and when multiple couplings are present simultaneously.   Collider signatures are explored in Section~\ref{sec:collider}.  To ascertain the current constraints, we present a phenomenological study of this signal based on a recast of a \acro{CMS} search for exotic multi-lepton final states~\cite{Khachatryan:2016iqn} in Section~\ref{sec:Multi-lepton:search:limits}.  We then design a dedicated search that would capitalize on the unique kinematics to greatly enhance sensitivity to the model in Section~\ref{sec:search}.  We comment on long-lived decays in Section~\ref{sec:LLP} and conclude in Section~\ref{sec:conclusions}.  We also provide several simple \acro{UV} completions for the low-energy theory in Appendix~\ref{app:UV:completion}, and review the chiral structure of these couplings in Appendix~\ref{app:chiral}.

%%%%%%%%%%%%%%%%%%%%%%%%%%%%%%%%%%%%
\section{A simplified model of lepton-flavor violating mediators}
\label{sec:model}
 
We focus on the flavor-violating couplings of charged leptons to a gauge singlet scalar mediator, $\varphi$. This mediator may be real or complex. In this section, we discuss the low-energy model and the electroweak gauge-invariant effective theory.  Appendix \ref{app:UV:completion} describes some simple \acro{UV} completions.  

%%%%%%%%%%%%%%%%%%%%%%%%%%%%%%%%%%%%
\subsection{Low-energy couplings to leptons}

For this study, we make the reasonable assumption that any additional hidden sector particles, e.g.~dark matter, do not influence the multi-lepton Higgs decays of interest.  This can be accomplished easily if, for instance, the other dark sector particles are heavier than the mediator. The interactions of $\varphi$ with charged leptons $\ell$ are encoded in the effective Lagrangian terms,
\begin{align}
  \mathcal{L}
    \supset 
    \left(y_{ij} \bar{\ell}_{i}P_L\ell_j\varphi 
            + y_{ij}^* \bar{\ell}_{j}P_R\ell_i\varphi^*\right)
    + \left(y'_{ij}\bar{\ell}_{i}P_R \ell_{j}\varphi 
            + y_{ij}'^* \bar{\ell}_{j}P_L \ell_{i}\varphi^*\right)
    \ ,
  \label{eq:fi:sm}
\end{align}
where $i,j$ index charged lepton flavor mass eigenstates ($1=e$, $2=\mu$, $3=\tau$). See Appendix \ref{app:chiral} for a brief review of the relevant chiral structure.
The terms within each parenthesis are related by complex conjugation. When $\varphi$ is real, the $y'$ terms are redundant and should be removed. 
Throughout most of this work, we assume that either $y_{ij}$ \emph{or} $y'_{ij}$ is non-zero for a specific pair of distinct flavors, $i\neq j$, and all other couplings are either exactly zero or negligibly small. For example, if we choose $y_{12}$ to be non-zero, then $y_{ij} = 0$ for all $i\neq 1$ and $j\neq 2$, and $y'_{ij} = 0$ for all $i$ and $j$. The absence of the diagonal coupling is relevant for the suppression of flavor constraints on the model.\footnote{Elements that we set to zero at tree-level are generated at loop-level by the charged lepton violation of the Standard Model, but these processes are suppressed by neutrino masses and are phenomenologically irrelevant to this study.}

The scalar $\varphi$ may also have a self-interaction potential.  These interactions are largely irrelevant for this study, with one notable exception: that we assume that $\varphi$ does not acquire a \acro{VEV}.  Such a \acro{VEV} would shift the Standard Model charged lepton mass eigenstates and introduce mixing.   In practice, a violation of the assumption would complicate the model, but for a sufficiently small \acro{VEV} the salient features of the exotic Higgs decays would be largely preserved and would not appreciably disrupt the behavior of the Higgs.   However, the misalignment of mass and Higgs interaction eigenstates could potentially generate sizable contributions to \acro{LFV} observables.

%%%%%%%%%%%%%%%%%%%%%%%%%%%%%%%%%%%%
\subsection{Gauge-invariant effective theory}

Due to the chiral nature of electroweak symmetry, the interactions of \eqref{eq:fi:sm} are not $SU(2)_L\times U(1)_Y$ gauge invariant. Those interactions must be generated by higher-dimension operators that include the Higgs at the electroweak scale.  The operators with the lowest possible dimension are 
\begin{align}
  \mathcal L_{\varphi\text{-lep.}}^{(\text{EW})} &= \frac{g_{ij}}{\Lambda}\,  \bar L_i H E_j\,  \varphi + \frac{g_{ij}'}{\Lambda}\,  \bar L_i H E_j\,  \varphi^* + \text{h.c.},
  &
\mbox{where } \; y_{ij}^{(')} &= \frac{g_{ij}^{(')}}{\Lambda} \frac{v}{\sqrt{2}},
  \label{eq:dim:5}
\end{align}
$H$ is the Higgs doublet, $L_i$, the lepton doublet of flavor $i$, and $E_j$, the lepton singlet of flavor $j$, are expressed in the mass eigenbasis, i.e.~the Standard Model Yukawas are diagonal, $\Lambda$ is a mass scale associated with the ultraviolet physics generating that these interactions, and the $g^{(')}$ are dimensionless effective couplings.
The low-energy couplings in \eqref{eq:fi:sm} are generated upon inserting the Higgs \acro{VEV}, $\langle H \rangle = (0 \, , \, v/\sqrt{2})$. In a case where $y_{ii} \neq 0$, these interactions generate tadpole term for $\varphi$ through the Higgs~\cite{Batell:2017kty}. 
Such a tadpole term would destabilize the $\varphi$ potential and necessarily generate a \acro{VEV}.  However, this two-loop tadpole is small enough that it would only minimally impact phenomenology.

Gauge invariance constrains the form of the scalar potential between  $\varphi$ and the Higgs:
\begin{align}
  V_{\varphi\text{-H}}
  &=
  \kappa H^\dagger H \varphi^*\varphi
  + \left( m_\varphi^2 -\kappa \frac{v^2}{2} \right) \varphi^*\varphi
  \label{eq:scalar}
  \, .
\end{align}
For the purposes of this study, we may ignore the quartic $\varphi$ self-interactions.
For real $\varphi$, there is an additional factor of one-half in \eqref{eq:scalar}, i.e.~$V_{\varphi\text{-H}} \to V_{\varphi\text{-H}} /2$. 

The underlying dynamics that generate the dimension-5 operators in (\ref{eq:dim:5}) are not relevant for the present study. As a proof of principle, we present three classes of renormalizable theories that generate (\ref{eq:dim:5}) above the electroweak scale in Appendix~\ref{app:UV:completion}: vector-like leptons, Froggatt--Nielsen fields, and $R$-parity violating supersymmetry. These do not require any additional light fields, so that it is consistent to examine the Higgs phenomenology of (\ref{eq:dim:5}) independently of a specific ultraviolet completion.

%%%%%%%%%%%%%%%%%%%%%%%%%%%%%%%%%%%%
\section{Flavor Constraints}
\label{sec:flavor}

%%%%%%%%%%%%%%%%%%%%%%%%%%%%%%%%%%%%

Spurious symmetries are a powerful tool in flavor physics~\cite{Gedalia:2010rj, Grossman:2017thq}. When $\varphi$ is complex, the interactions respect a spurious global $L_i-L_j$ symmetry under which the mediator is charged, $[\varphi] = -2$. This symmetry prohibits many charged lepton-flavor violating tree-level processes~\cite{Bernstein:2013hba}.
The symmetry is explicitly broken by interactions with the $W$-boson so that higher-order flavor violating processes are suppressed by a loop factor and the ratio of the neutrino and $W$-boson masses; see Appendix~\ref{app:chiral}. 
When $\varphi$ is real, there is no such spurious $L_i - L_j$ symmetry preserved by the interactions. In other words, a real field cannot carry a $U(1)$ charge. 

In this section, we first examine the flavor constraints where only the single, off-diagonal coupling is non-zero.  We then explore the flavor constraints in cases with multiple non-zero couplings.

%%%%%%%%%%%%%%%%%%%%%%%%%%%%%%%%%%%%
\subsection{Pure off-diagonal, single coupling}
\label{sec:pure:off:diagonal}

The limit where the $\varphi$ has only a single off-diagonal flavor coupling is protected from charged lepton-flavor violating processes, insulating the model against most constraints. We briefly review the relevant constraints and refer to Section~5.4 of Ref.~\cite{Galon:2016bka} for an in-depth discussion. 

A mediator with purely \acro{LFV} interaction can mediate potentially dangerous tree-level charged flavor violation in one notable case: the muonium system, a bound state of an electron and an anti-muon.  When $\varphi$ is real and couples to muons and electrons, it may mediate muonium--anti-muonium oscillations~\cite{Willmann:1998gd}. By recasting bounds on $R$-parity violating sneutrinos~\cite{Kim:1997rr}, one obtains a strong bound, 
\begin{align}
  y_{21} \, ,\; y_{12} &< 4.4 \times 10^{-4} \left(\frac{m_\varphi}{\text{GeV}}\right)\quad \left[\text{90\% \acro{CL}}\right]
  &
  \text{($\varphi$ real)} \ .
  \label{eq:muonium:bound}
\end{align}
When $\varphi$ is complex, the $L_{\mu} - L_{e}$ spurious symmetry prevents this process at tree-level. 

A bound that applies for both real and complex $\varphi$ comes from the interference of $t$-channel $\varphi$ exchange to the forward--backward asymmetry of $e^+e^-\to f\bar f$ scattering~\cite{langacker1996precision},
\begin{align}
  A_\text{FB}^f 
  &= 
  \frac{\sigma_>(e^+e^-\to f\bar f) - \sigma_<(e^+e^-\to f\bar f)}{\sigma_>(e^+e^-\to f\bar f) + \sigma_<(e^+e^-\to f\bar f)} \ .
\end{align}
Unlike the coupling constraints for muonium oscillation, these bounds hold for real or complex $\varphi$.
One may recast the $A_\text{FB}^f$ bounds on sneutrino interactions~\cite{Barger:1989rk, Barbier:2004ez} to the case of a lepton-flavor violating mediator, which yields
\begin{align}
  y_{21}^{\left(\prime\right)} \, ,\; y_{12}^{\left(\prime\right)}  & < 2.5\times 10^{-3} \left(\frac{m_\varphi}{\text{GeV}}\right)
  &
  y_{31}^{\left(\prime\right)} \, ,\; y_{13}^{\left(\prime\right)}  & < 1.1\times 10^{-3} \left(\frac{m_\varphi}{\text{GeV}}\right)
  &
  \left[\text{95\% \acro{CL}}\right] \ ,
  \label{eq:bound:FB}
\end{align}
while there are no constraints for  $y_{23}^{\left(\prime\right)}$, $y_{32}^{\left(\prime\right)}$.  

The spurious flavor symmetry also prevents tree-level contributes to rare multi-body lepton decays, $\ell_i \to \ell_j \ell_k \bar\ell_k$ and $\ell_i \to \ell_j \ell_k\bar\ell_k\nu\bar\nu$. When $\varphi$ is real, this process is still prohibited at tree-level because a $\varphi$ emitted by the initial heavy lepton must yield another lepton of the same heavy flavor. In other words, $\varphi$ does not contribute at tree-level to these decays in the absence of a flavor-diagonal coupling to the lighter lepton.  Flavor-changing dipole operators in this scenario are down by a loop factor from the Standard Model contribution that is itself suppressed by neutrino masses.   As $\varphi$ is hadrophobic, $\mu\to e$ conversion in the presence of nuclei only occurs at loop-level and, again, requires neutrino masses to violate flavor in the loop. 

In summary, the models with a real $\varphi$ with $e$ and $\mu$ are constrained by muonium oscillations \eqref{eq:muonium:bound}. For either real or complex $\varphi$ interacting with electrons, the primary single-coupling bound comes from precision measurements of lepton forward--backward asymmetries at lepton colliders \eqref{eq:bound:FB}. Otherwise, the single-coupling limit of this model produces negligible signals at traditional charged lepton-flavor violation experiments.  There are no appreciable constraints on $\varphi$ coupling with just the $\mu$ and $\tau$.

%%%%%%%%%%%%%%%%%%%%%%%%%%%%%%%%%%%%
\subsection{Multi-coupling}

Section~\ref{sec:pure:off:diagonal} shows that precision flavor constraints only moderately impact the single, off-diagonal coupling limit of the $\varphi$. For a more general flavor structure, additional lepton flavor-violating  constraints can enter.  Given that our scalar is assumed to connect only to the leptonic sector, the primary observables that constrain the model are the exotic decays $\ell_i\to\ell_j\gamma$ and $\ell_i\to \ell_j\ell_k\ell_l$.  To estimate the effect of relaxing the single-coupling assumption, we examine the effect of turning on an additional coupling.  
 
  \begin{figure}[t]
  \centering
 \includegraphics[width=0.49\textwidth]{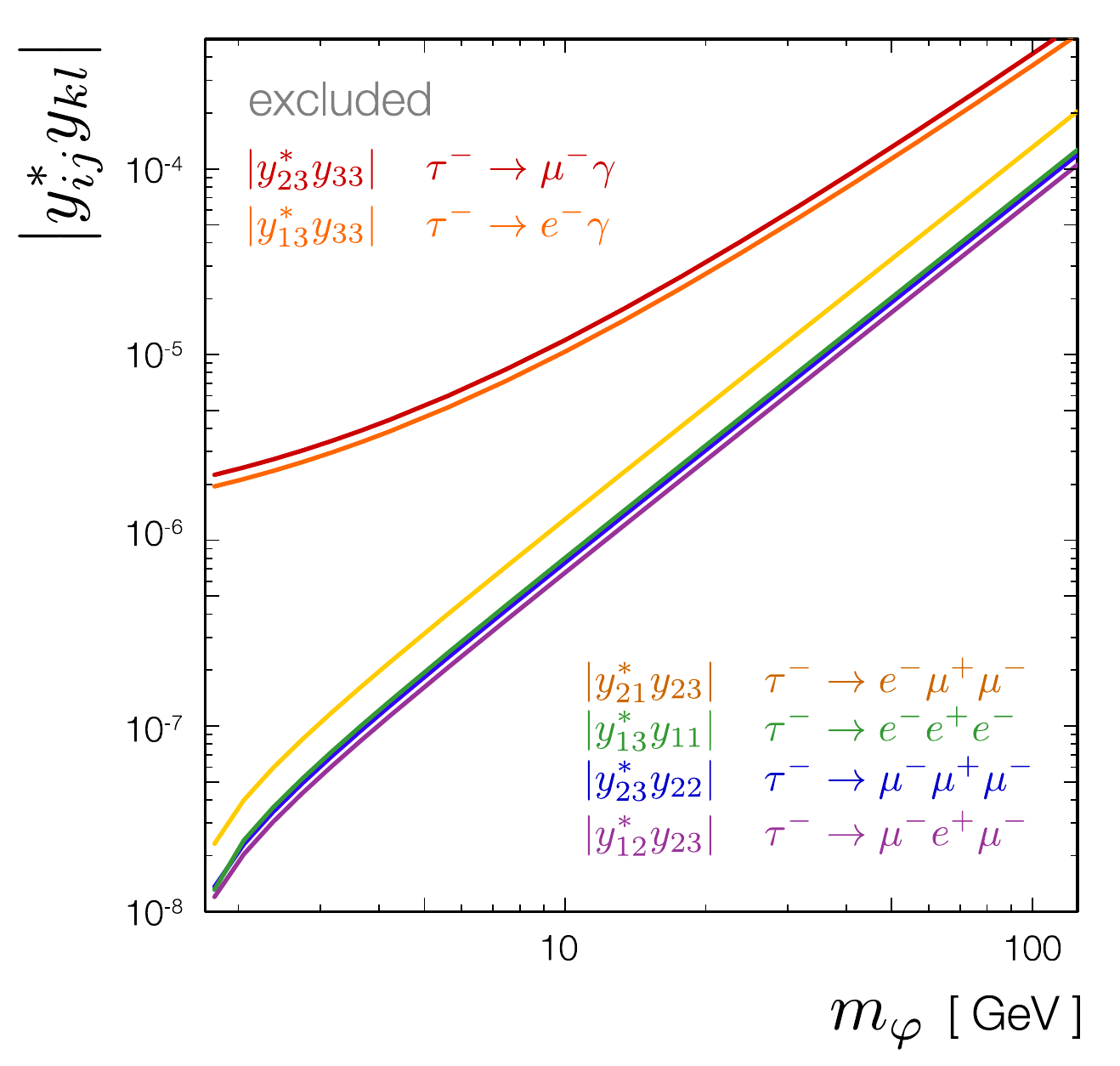}
 \hfill
 \includegraphics[width=0.49\textwidth]{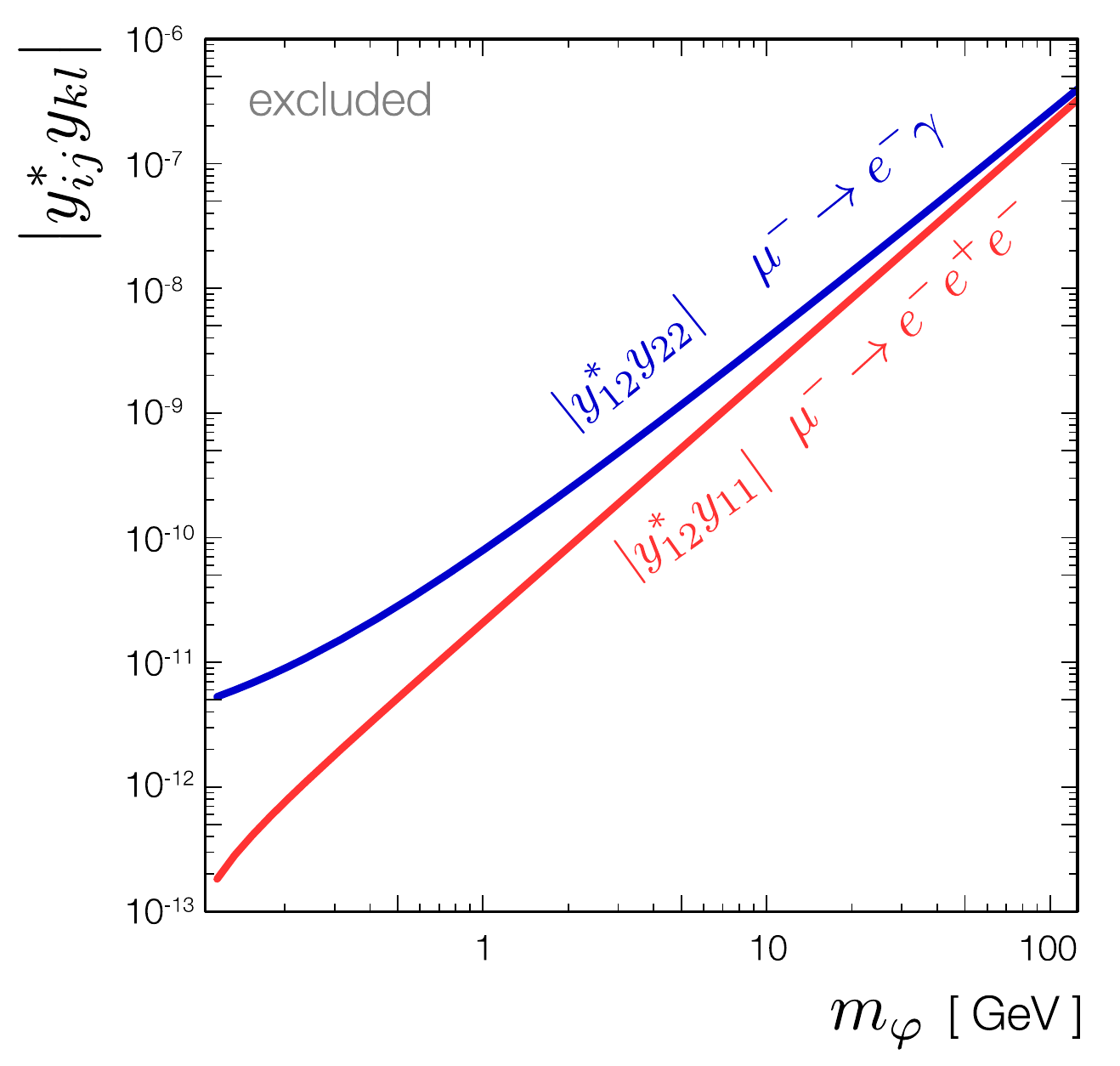}
 \caption{\label{fig:flavor}  Constraints on the coupling $\abs{y_{ij}^*y_{kl}}$ as a function of mass.  \textbf{Left: } \acro{LFV} constraints from $\tau$ decays.   \textbf{Right: } \acro{LFV} constraints from $\mu$ decays. }
\end{figure}

One of the most promising channels to observe lepton flavor violation is the radiative decays of leptons.  The \acro{LFV} radiative decay branching ratio for a lepton is \cite{Harnik:2012pb}
\beq
\text{Br}(\ell_i\to\ell_j\gamma) =\frac{1}{\Gamma_i}\frac{\alpha_{em}m_i^5}{64\pi^4}\lp\abs{C_L}^2+\abs{C_R}^2\rp,
\label{eq:BRrad}
\eeq
where, in this model, the Wilson coefficients can be expressed as
\begin{align}
C_L &= \sum_k F(m_i,m_k,m_j, m_\varphi,y) 
&\text{ and }&&
C_R &= \sum_k F(m_i,m_k,m_j, m_\varphi,y^\dagger) 
\end{align}
for the loop function 
\beq
F(m_i,m_k,m_j, m_\varphi,y) = \int^1_0 dxdydz\delta(1\!-\!x\!-\!y\!-\!z) \frac{xzm_j y_{jk}y^*_{ik}+yzm_i y^*_{kj}y_{ki}+(x+y)m_k y^*_{kj}y^*_{ik}}{4m_i\lp zm_\varphi^2 -xzm_j^2 -yzm_i^2 +(x+y)m_k^2\rp}.
\label{eq:looprad}
\eeq
The measured bounds on these processes for the $\tau$ decays are Br$(\tau\to \mu \gamma)<4.4\times 10^{-8}$ and Br$(\tau\to e \gamma)<3.3\times 10^{-8}$ from  BaBar \cite{Aubert:2009ag}, and Br$(\mu\to e \gamma)<4.2\times 10^{-13}$ from the \acro{MEG} experiment \cite{TheMEG:2016wtm}.  In the near future, Belle~\acro{II}  \cite{Kou:2018nap} and \acro{MEG}-\acro{II} \cite{Baldini:2018nnn}  are expected to improve on these constraints.

 The decay of a lepton into three lighter leptons, $\ell_i^-\to \ell_j^-\ell_k^+\ell_l^-$, can be facilitated through tree-level $\varphi$ exchange.  The amplitude for these decays can be written 
 \beq
\mathcal M_{ijkl} = \frac{\left[\bar u(p_i) \lp y^*_{ij} P_L +y_{ji} P_R \rp u(p_j)\right]\left[\bar v(p_k) \lp y^*_{kl} P_L +y_{lk} P_R \rp u(p_l)\right]}{m_{kl}^2 -m_\varphi^2 + im_\varphi\Gamma_\varphi} + \delta_{jl} \left\{ j\leftrightarrow l \right\} \label{eq:amp3l}
 \eeq
 with the invariant masses $m_{ij}^2 \equiv (p_i +p_j)^2$.  After squaring and spin averaging, the final branching fraction into three leptons can be computed as
 \beq
\text{Br}(\ell_i^-\to\ell_j^-\ell_k^+\ell_l^-) =\frac{1}{ \Gamma_i} \int \frac{\abs{\mathcal M_{ijkl}}^2}{512 \pi^3 m_i^3} d m_{jk}^2d m_{kl}^2.
\label{eq:BR3l}
\eeq
Belle places the most stringent constraints on all $\tau^-\to\ell_j^-\ell_k^+\ell_l^-$ branching fractions~\cite{Hayasaka:2010np}, while the $\mu^-\to e^-e^+e^-$ bound is from \acro{SINDRUM}~\cite{Bellgardt:1987du}:  
\begin{align*}
\text{Br}(\tau^- \to \mu^-\mu^+\mu^-)&<2.1\times 10^{-8}, &
\text{Br}(\tau^- \to \mu^-\mu^+e^-)&<2.7\times 10^{-8}, \\
\text{Br}(\tau^- \to \mu^-e^+\mu^-)&<1.7\times 10^{-8}, &
\text{Br}(\tau^- \to \mu^-e^+e^-)&<1.8\times 10^{-8}, \\
\text{Br}(\tau^- \to e^-\mu^+e^-)&<1.5\times 10^{-8}, &
\text{Br}(\tau^- \to e^-e^+e^-)&<2.7\times 10^{-8}, \\
& \text{and} &
\text{Br}(\mu^- \to e^-e^+e^-)&<1.0\times 10^{-12}. 
\end{align*}
Belle~\acro{II}~\cite{Kou:2018nap} and Mu3e~\cite{Perrevoort:2016nuv} are expected to improve these limits considerably.

The most stringent constraint for each combination of two couplings is shown in Fig.~\ref{fig:flavor}.  Restricting to two of nine couplings in $y$, allows us to directly bound $\abs{y_{ij}^*y_{kl}}$.\footnote{If $y_{ii}$ is complex, the contributions in \eqref{eq:looprad} depend on the phase of $y_{ii}$.  However, it is difficult to induce qualitative changes to limits by modifications to this phase.  We present real $y_{ii}$ in expressing the limits.}  In comparing these constraints, one should consider that $\abs{y_{ij}^*y_{kl}}$ is constrained here, whereas in the Section~\ref{sec:pure:off:diagonal}, the limits are presented on a single power of $y_{ij}$.  Most of the flavor constraints admit a large parameter space that allows for prompt decays.  Unsurprisingly, coupling combinations that involve only the first and second generation are more tightly constrained.

%%%%%%%%%%%%%%%%%%%%%%%%%%%%%%%%%%%%
\section{Collider}
\label{sec:collider}

In this section, we explore the collider sensitivity to the \acro{LFV} scalar model.
We recast a multi-lepton search by the \acro{CMS} experiment, propose a dedicated search, and comment briefly on the model when $\varphi$ is a long-lived particle.

%------------------------------------
\begin{figure}[t]
 \includegraphics[width=0.49\textwidth]{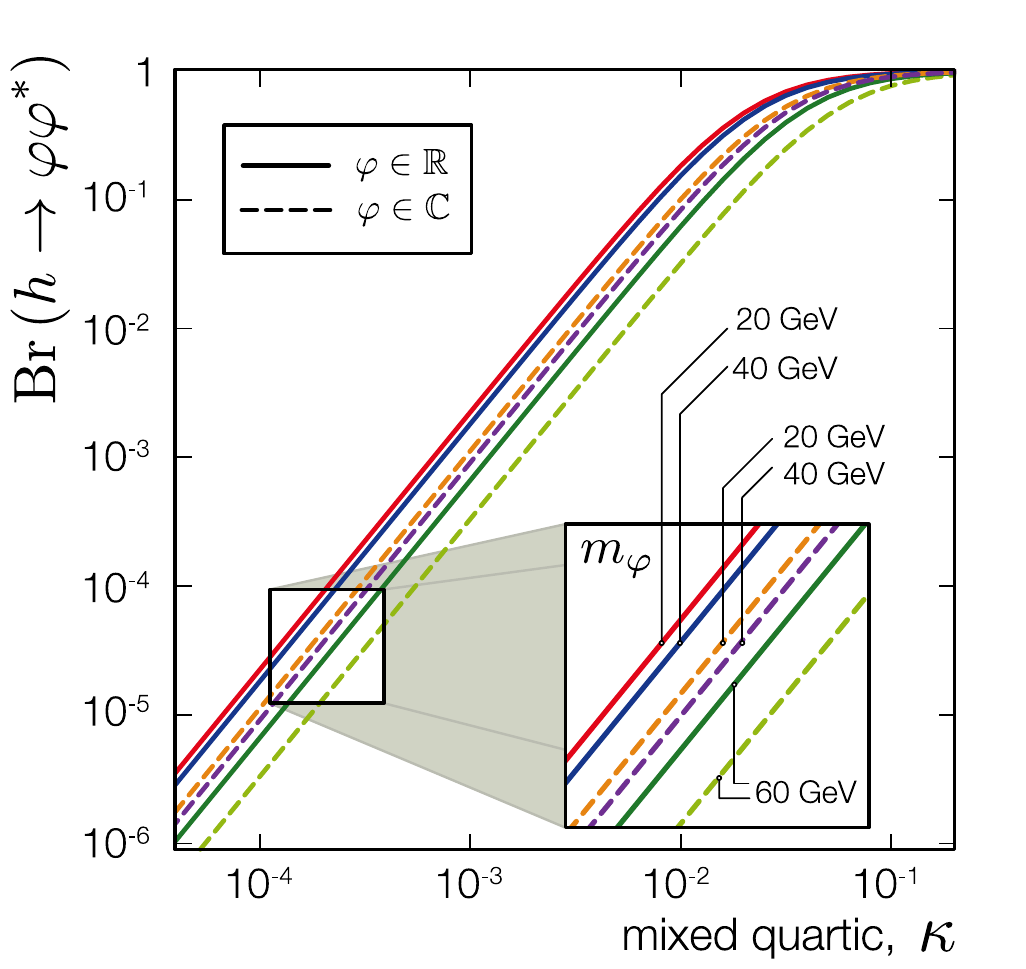}
 \hfill
 \includegraphics[width=0.49\textwidth]{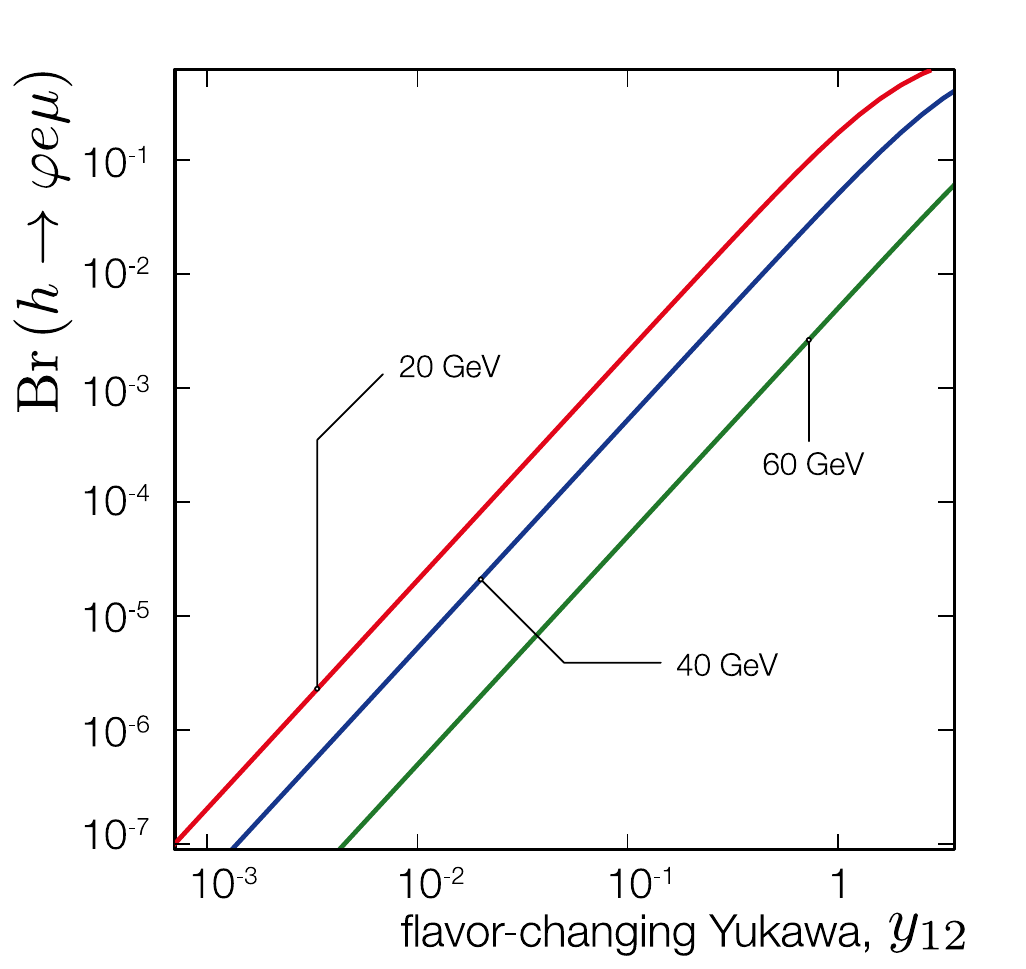}
 \caption{\label{higgs:br}The branching ratio of \textbf{Left:} $h \to \varphi\varphi^*$
  and \textbf{Right: } $h \to \varphi\, e^- \mu^+$ processes as a function of $\kappa$ (left) and
   $y_{12}$ (right) for various masses of $\varphi$. The dashed curves are for the cases with real $\varphi$.}
   \label{fig:BRinvars}
\end{figure}
%------------------------------------

The model outlined in Section~\ref{sec:model} allows for two possible exotic Higgs decays into the \acro{LFV} scalar.  The pertinent terms of the Lagrangian are
\beq
\mathcal L \sim \lp 
\frac{y_{ij}}{v/\sqrt{2}}\,  \bar L_i H E_j\,  \varphi +\hc \rp +  \kappa H^\dagger H \varphi^*\varphi.
\label{eq:L}
\eeq
The first term in \eqref{eq:L} introduces both the decay, $\varphi\to\ell_i^+\ell_j^-$, and the Higgs decay mode, $h\to \varphi \ell_i^+ \ell_j^-$.  Despite the minimality of this option, a more compelling pathway comes from the marginal operator of the second term. This generates the $h\to \varphi\varphi^*$ decay. The Higgs decay widths for these processes are
\begin{align}
  \label{eq:h:width}
\Gamma_{h\to \varphi \ell_i^+ \ell_j^-} 
  =& \frac{y_{ij}^2}{128 \pi^3 m_h^3 v^2}  \int \left[ m_{ij}^2 - (m_{i} + m_{j})^2 \right] d m_{\varphi i}^2\, d m_{ij}^2 % \, 
  \\
\Gamma_{h\to \varphi \varphi^*} =& \frac{\kappa^2 v^2}{16 \pi m_h} \sqrt{1 - \left(\frac{2m_{\varphi}}{m_h}\right)^2}
\ ,
\end{align}
where there is an extra factor of $1/2$ in the second line for the case with real $\varphi$. Fig.~\ref{fig:BRinvars} shows the branching ratio of the  Higgs with Standard Model-like partial width into $h\to \varphi\varphi^*$ and $h\to \varphi \ell_i^+ \ell_j^-$  as a function of the couplings in \eqref{eq:L} for three choices of $m_\varphi$.  

The first term in \eqref{eq:L} permits the decay of $\varphi$ into two charged leptons of different flavors.  When $\varphi$ is complex, the spurious $L_i-L_j$ symmetry prevents the Higgs decays from producing same-sign lepton pairs, e.g.~$\varphi \to e^+\mu^-$ and $\varphi^* \to e^-\mu^+$. 
On the other hand, when $\varphi$ is real, there is no such symmetry and each $\varphi$ decays with equal probability into either sign, e.g.~$\varphi\to e^+\mu^-$ or $e^-\mu^+$.  Thus, a real $\varphi$ means that Higgs decays can yield exotic same-sign lepton pairs.

In the pure off-diagonal, single-coupling case, we explore four different models based on whether $\varphi$ is a real or complex scalar, and on what coupling $y_{12}$ or $y_{23}$ is non-zero.  We expect the collider phenomenology of the $y_{13}$ case to be very similar to the $y_{23}$ case, and do not study it in further detail here, but we note that \eqref{eq:bound:FB} bounds $y_{13}$.  We list the final states generated for each scenario in Table~\ref{tab:models}.  Throughout this section, we assume the off-diagonal Yukawa coupling that enables the $\varphi$ decay is large enough for a prompt signature.  This requires $y_{ij} \gtrsim 10^{-6}$.  This implies that the new physics generating the dimension-5 operators may not be to far above the PeV scale.   

\begin{table}[t]
 \centering
  \begin{tabular}{@{}l@{\hskip 2em}l@{\hskip 2em}l@{\hskip 2em}l@{}}
  \toprule 
  Model & $\varphi$ & Coupling & Final States \\
  \midrule
  \ModelI & $\mathbb{C}$ & $y_{12}\neq0$ & $e^+ e^- \mu^+ \mu^-$ \\
  \ModelII & $\mathbb{C}$ & $y_{23}\neq0$ & $\mu^+ \mu^- \tau^+ \tau^-$ \\
  \ModelIII & $\mathbb{R}$ & $y_{12}\neq0$ & $e^+ e^+ \mu^- \mu^-$ \\
  \ModelIV & $\mathbb{R}$ & $y_{23}\neq0$  & $\mu^+ \mu^+ \tau^- \tau^-$ \\
  \bottomrule
  \end{tabular}
 \caption{The four pure off-diagonal, single-coupling models considered in this section.  In the cases with real $\varphi$, the opposite-sign cases appear with equal frequency to the more striking same-sign cases listed. We expect $y_{13}\neq 0$ models to have very similar exotic Higgs decay phenomenology to the analogous $y_{23}\neq 0$ model.}
 \label{tab:models}
\end{table}

Even when the \acro{LFV} couplings are small, they could potentially generate an additional \acro{LFV} signature from $\varphi$-strahlung off of produced lepton pairs.  Notably, one could have the decay $Z \to  \varphi \ell_i^+ \ell_j^-$.  Unless $\varphi$ is very light, the $Z$ is the only potentially sizable production pathway beyond Higgs decays.  As we will illustrate in Section~\ref{sec:Multi-lepton:search:limits}, $Z$ production is always subdominant to the Higgs pathways.

%%%%%%%%%%%%%%%%%%%%%%%%%%%%%%%%%%%%
\subsection{Multi-lepton search limits}
\label{sec:Multi-lepton:search:limits}

We present the limits from a recasted search for multi-lepton signals by the \acro{CMS} experiment~\cite{Khachatryan:2016iqn} corresponding to an integrated luminosity of 19.5~fb$^{-1}$ at $\sqrt{s}= 8$ TeV.
We consider this specific search as it contains bins with no missing energy requirement, arbitrarily low $S_T$, and fairly soft cuts on lepton $p_T$.  The overwhelming majority of available multi-lepton searches fail to satisfy one or more of these conditions \cite{Khachatryan:2017qgo,Sirunyan:2017lae,Sirunyan:2018ubx,Aaboud:2018zeb,Sirunyan:2019ofn}.  There are also searches for a Higgs decaying into four leptons via two pseudoscalars performed at the Tevatron~\cite{Abazov:2009yi} and the \acro{LHC}~\cite{Aad:2015oqa, Khachatryan:2017mnf,Aaboud:2018fvk}. However, in each of those searches the new particle is assumed to decay into same-flavor lepton pairs. 

We simulated the production of multi-lepton final states in our model assuming the same event selection criteria as Ref.~\cite{Khachatryan:2016iqn}. The model is implemented in {\tt FeynRules}~\cite{Alloul:2013bka,Christensen:2008py}, and we use {\tt MadGraph-5}~\cite{Alwall:2014hca} to generate the parton level events. These events are fed into {\tt PYTHIA-8}~\cite{Sjostrand:2014zea} for showering and hadronization using {\tt CTEQ6L1} as the parton distribution function and {\tt FastJet-3}~\cite{Cacciari:2011ma} for jet reconstruction. The program used for the collider analysis is available on GitHub.\footnote{\texttt{\href{https://github.com/ZAKI1905/Pheno}{https://github.com/ZAKI1905/Pheno}}}

\begin{figure}[t]
 \begin{center}
 \includegraphics[width=\textwidth]{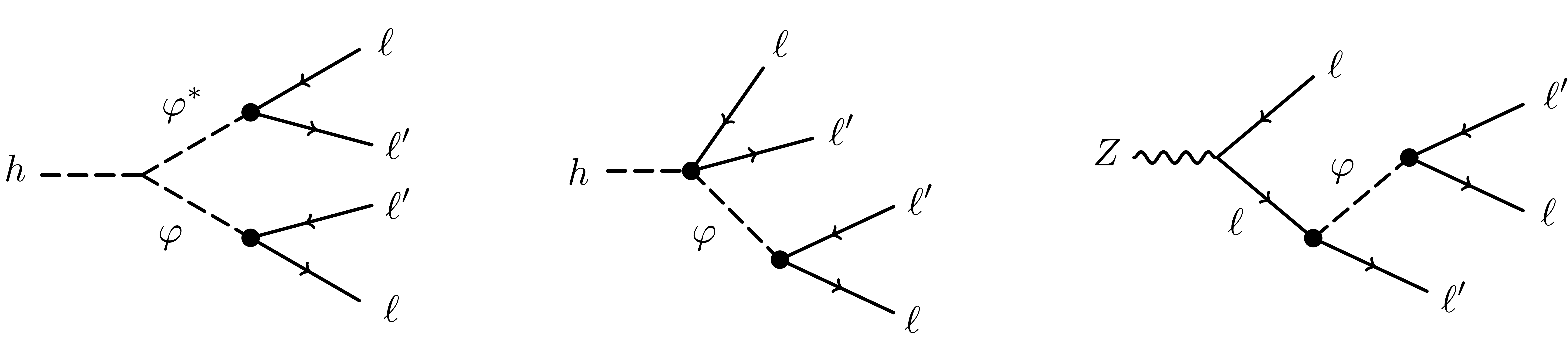}
 \end{center}
 \caption{Feynman diagrams for the processes listed in Table~\ref{tab:process}: 
  The dominant diagram for the case with a non-zero quartic coupling $\kappa\neq 0$ (left), 
  the main decay channels involving the Higgs (middle), and $Z$ boson 
  (right) in the case with $\kappa= 0$. }
 \label{fig:diags}
\end{figure}

\begin{figure}[t]
 \centering
 \includegraphics[width=.49\textwidth]{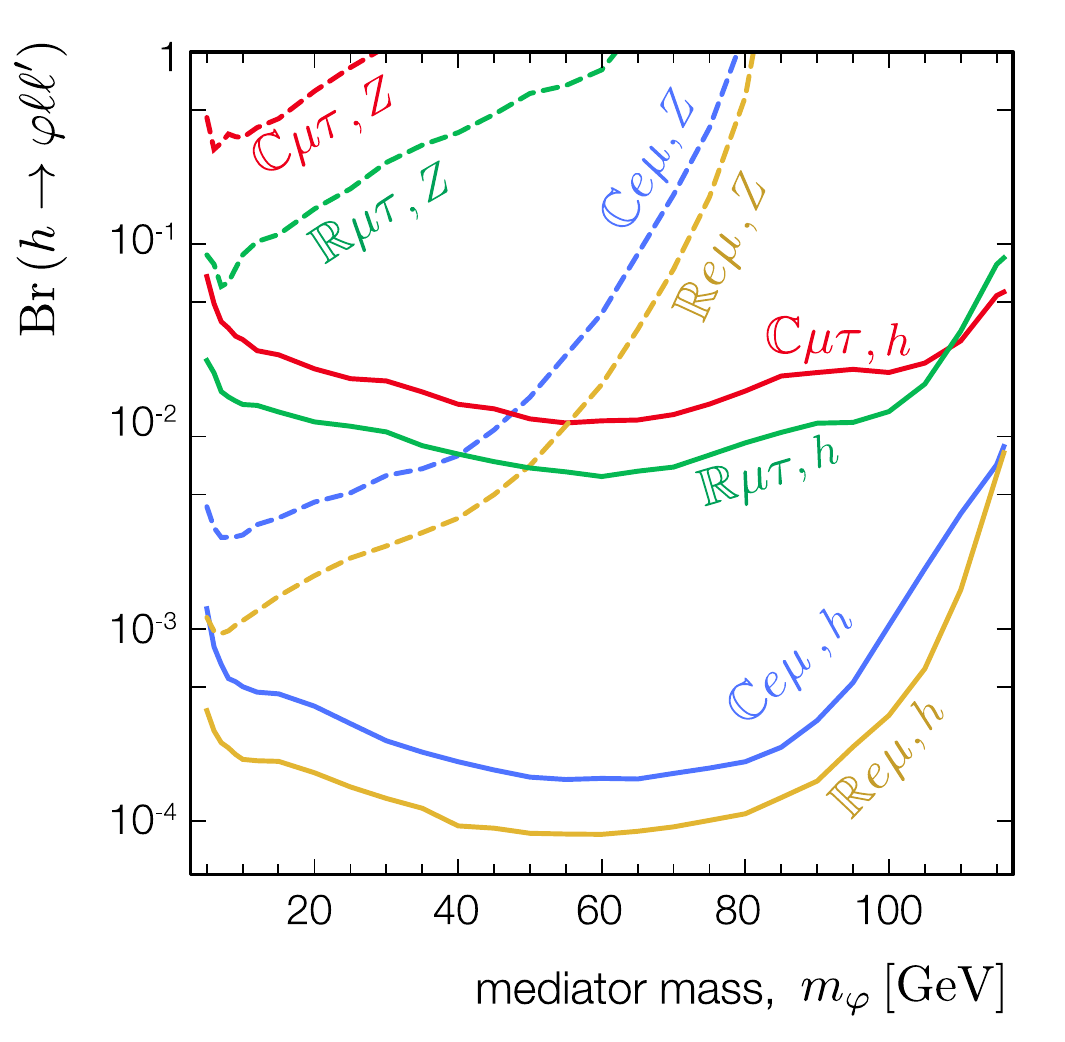}
 \hfill
 \includegraphics[width=.49\textwidth]{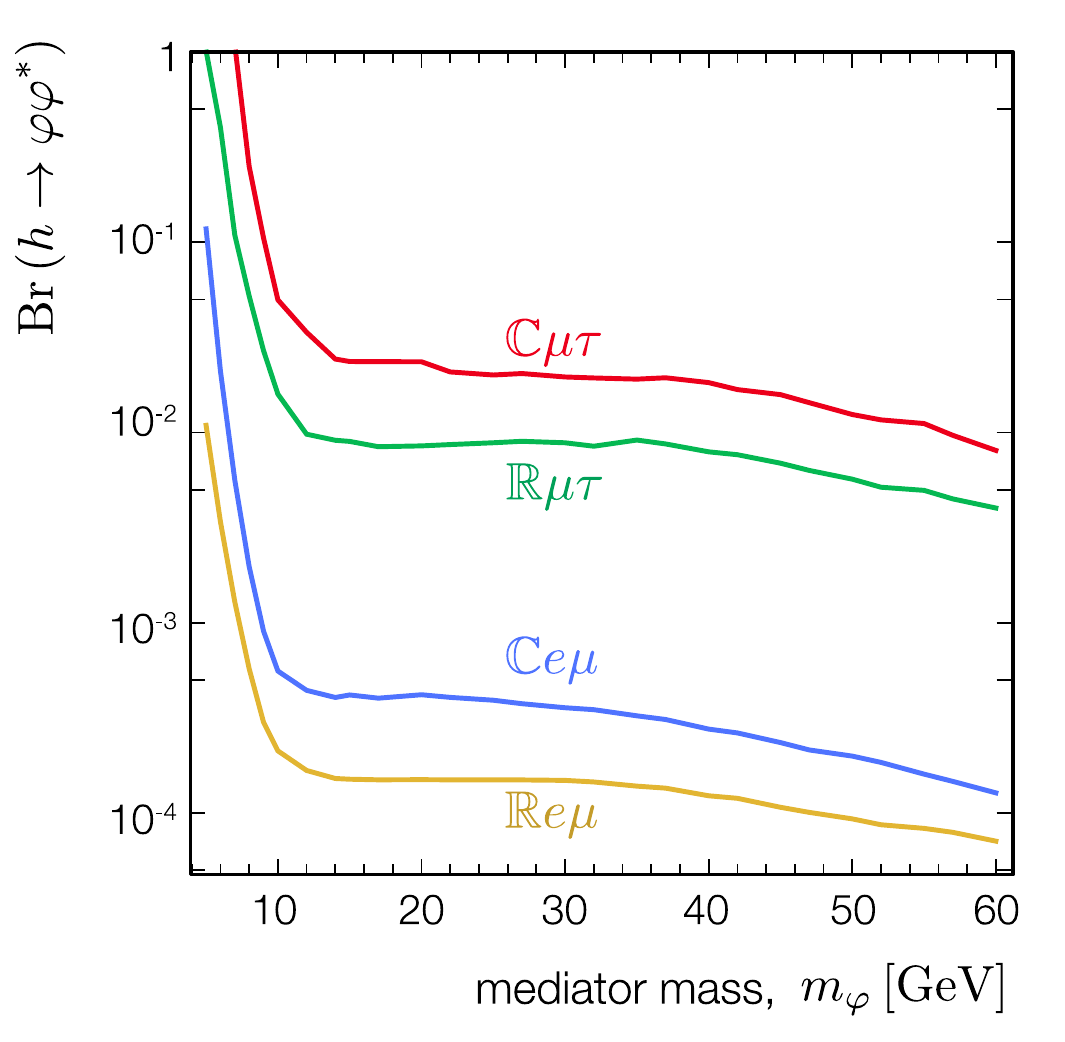}
 \caption{Limits on the branching ratio at 95\%~\acro{CL} for the processes listed in Table~\ref{tab:process}. \textbf{Left:} The limits on the branching ratio from the Higgs-mediated channels (\textbf{solid}) in the case with $\kappa= 0$. The limits on $Z$-mediated channels are translated into the equivalent limits on Br$(h \to \varphi \ell \ell')$ using their analytical  expressions in~\ref{eq:z2h} (\textbf{dashed}).   \textbf{Right:} The limits for the case with a non-zero quartic coupling $\kappa\neq 0$. }
 \label{fig:CMS:limits}
\end{figure}

We consider four-lepton production through three different processes for each of the models as listed in Table~\ref{tab:process} (see Fig.~\ref{fig:diags}).  In addition to the same-sign process shown, the cases with real $\varphi$ have the opposite-sign final states included when we derive the bounds in Fig.~\ref{fig:CMS:limits}. 

\begin{table}[t]
\centering
\begin{tabular}{@{}l@{\hskip 2em}l@{\hskip 2em}lllll@{\hskip 2em}r@{\hskip .2em}l@{}}
\toprule
Model & Quartic &\multicolumn{5}{l}{\hspace{-0.5em}Process ($pp$ collider)} & \multicolumn{2}{l}{Range of $m_\varphi$}
\\
\midrule
\multirow{3}{*}{\ModelI, \ModelII}
  & $\kappa \neq 0$ 
    & $h$ 
    & $\to$ 
    & $\varphi + \varphi^*$ 
    & $\to$ 
    & $\ell_i^+ + \ell_i^- + \ell_j^+  + \ell_j^-$ 
    & [5 -- 60]&GeV
    \\
  & $\kappa = 0$ 
    & $h$  
    & $\to$  
    & $\varphi + \ell^+_i + \ell_j^-$ 
    & $\to$  
    & $\ell_i^+ + \ell_i^- + \ell_j^+  + \ell_j^-$  
    & [5 -- 120]&GeV 
    \\
  & $\kappa = 0$ 
    & $Z$  
    & $\to$  
    & $\varphi + \ell^+_i + \ell_j^-$ 
    & $\to$  
    & $\ell_i^+ + \ell_i^- + \ell_j^+  + \ell_j^-$  
    & [5 -- 90]&GeV 
  \\
\midrule
\multirow{3}{*}{\ModelIII, \ModelIV}
  & $\kappa \neq 0$ 
    & $h$ 
    & $\to$ 
    & $\varphi + \varphi$ 
    & $\to$ 
    & $\ell_i^+ + \ell_i^+ + \ell_j^-  + \ell_j^-$ 
    & [5 -- 60]&GeV 
    \\
  & $\kappa = 0$ 
    & $h$  
    & $\to$  
    & $\varphi + \ell^+_i + \ell_j^-$  
    & $\to$  
    & $\ell_i^+ + \ell_i^+ + \ell_j^-  + \ell_j^-$  
    & [5 -- 120]&GeV 
    \\
  & $\kappa = 0$ 
    & $Z$ 
    & $\to$  
    & $\varphi + \ell^+_i + \ell_j^-$  
    & $\to$  
    & $\ell_i^+ + \ell_i^+ + \ell_j^-  + \ell_j^-$  
    & [5 -- 90]&GeV 
  \\
  \bottomrule
\end{tabular}
 \caption{\label{tab:process}The different channels considered in our analysis, along with the ranges of the masses for $\varphi$. For $\kappa=0$, we ignore the interference between the $h$- and $Z$-mediated channels.}
\end{table}

A summary of the cuts implemented from Ref.~\cite{Khachatryan:2016iqn} are presented in Table~\ref{tab:cuts:cms}.  Lepton identification is modeled using the quoted \acro{CMS} \acro{ID} efficiencies in Ref.~\cite{CMS:2013qda} for electrons and muons, and assuming a $70\%$ efficiency for the hadronic taus~\cite{Khachatryan:2016iqn}. The momenta of leptons are corrected using the \acro{CMS} detector resolution for electron momentum~\cite{Khachatryan:2015hwa} and assuming a resolution of 1\% in the barrel ($|\eta| < 0.9$), 3\% in the endcaps ($1.2 \leq |\eta| \leq 2.4$), and 2\% in the overlap region ($0.9 \leq |\eta| < 1.2$) for muons~\cite{Sirunyan:2018fpa}. The events that pass the cuts are then classified based on the number of leptons ($N_{\ell}$), hadronically decaying taus ($N_{\tau_h}$), and the number of opposite-sign same-flavor leptons ($N_{\text{OSSF}}$). The events are further binned according to their $S_T$ values, which is the sum of the missing transverse momentum and the scalar sum of the $p_T$ of all  jets and charged leptons. If there exists an \acro{OSSF} lepton pair  for which $75 \text{ GeV } < m_{{\ell}^+{\ell}^-} < 105 \text{ GeV}$, the events are categorized as on-$Z$, otherwise they are labeled off-$Z$.  When $N_{\ell}=3$, the off-$Z$ region is additionally divided into $m_{{\ell}^+{\ell}^-} < 75$~GeV and 105~GeV $< m_{{\ell}^+{\ell}^-}$.

\begin{table}[H]
 \centering
 \begin{tabular}{l@{\hskip 3em}rcl}
 \toprule
 Objects & \multicolumn{3}{c}{Conditions} \\
 \midrule
 $e^+e^-,\, \mu^+\mu^-$ 
 & 
  $m_{\ell^+\ell^-}$
  & $>$ &
  12~GeV
  \\
\midrule
\multirow{4}{*}{$e,\, \mu$}
& 
  $p_T^\text{lead}$
  & $>$ &
  20~GeV \vspace{-2.5mm}
  \\
& 
  $p_T^\text{sub}$
  & $>$ &
  10~GeV \vspace{-2.5mm}
  \\
& 
  $|\eta|$
  & $<$ &
  2.4 \vspace{-2.5mm}
  \\
& 
  $\sum\limits_{R({\ell}, i) < 0.3} \, E_T^i$
  & $<$ &
  2~GeV 
    \\
\midrule
\multirow{3}{*}{$\tau_h$}
& 
  $p_T$
  & $>$ &
  20~GeV \vspace{-2.5mm}
  \\
& 
  $|\eta|$
  & $<$ &
  2.3 \vspace{-2.5mm}
  \\
& 
  $\sum\limits_{0.1<R(\tau_h,i)<0.3} \, E_T^i$
  & $<$ &
  2~GeV 
  \\
\midrule
\multirow{3}{*}{jets}
& 
  $p_T$
  & $>$ &
  30~GeV \vspace{-2.5mm}
  \\
& 
  $|\eta|$
  & $<$ &
  2.5 \vspace{-2.5mm}
  \\
& 
  $\Delta R(\text{jet}, \ell)$
  & $>$ &
  0.3 \vspace{-1mm}
  \\
 \bottomrule
 \end{tabular}
 \caption{\label{tab:cuts:cms}Summary of the cuts implemented in our simulation~\cite{Khachatryan:2016iqn}. Jets are reconstructed using the anti-kT algorithm~\cite{Cacciari:2008gp} with a distance parameter $R = 0.5$, and $\ell$ in the last row stands for the isolated electron, muon, or $\tau_h$ candidates.}
\end{table}

In order to use the \acro{CMS} results to constrain our models, we apply the following procedure.  In each bin, we use a Poisson distribution for $n$ events given an expected rate of $\epsilon \cdot \mathcal{L} \cdot \sigma + B$:
\beq
P(n\, ; \epsilon \cdot \mathcal{L} \cdot \sigma + B) = \frac{e^{-(\epsilon \cdot \mathcal{L} \cdot \sigma + B)}}{n!} (\epsilon \cdot \mathcal{L} \cdot \sigma + B)^n\, 
\eeq
where $\epsilon$ and $B$ are the signal efficiency and the expected background in that specific bin respectively, $\mathcal{L} = 19.5$ fb$^{-1}$ is the integrated luminosity, and
$\sigma$ is the cross-section for the process under consideration. This cross-section can be written as
\beq
\label{eq:prodxsec}
\sigma = \sigma( pp \to X) \cdot \text{ Br}( X \to 4l) \qquad\qquad  X= h,Z
\eeq 
 where we use the 8~TeV cross-sections $\sigma( p p \to h) = 24.2$~pb~\cite{PDG:Higgs:2018},  $\sigma( p p \to Z) = 34.17$~nb~\cite{Chatrchyan:2014mua}, and Br is the branching ratio on which we want to set the limit. The total likelihood is
\beq
\label{eq:likelihood}
L =  \prod\limits_{i} \frac{e^{-(\epsilon_i \cdot \mathcal{L} \cdot \sigma + B_i)}}{n_i!} (\epsilon_i \cdot \mathcal{L} \cdot \sigma + B_i)^{n_i} \cdot P_{\text{LN}}(B_i | \bar{B}_i, \delta B_i ) \ ,
\eeq
where the product is over the bins. We model the systematic uncertainty associated with the background prediction as log-normal functions $P_{\text{LN}}(B_i | \bar{B}_i, \delta B_i )$ for the measured value $B_i$, which depends on the expected value $\bar{B}_i$ and an uncertainty $\delta B_i$.  We use the hybrid Bayesian--frequentist approach~\cite{COUSINS1992331} to marginalize the nuisance parameters by integrating over the background ($B_i$) errors. We then use this marginal likelihood ($L_m$) to form the log-likelihood ratio test statistic defined as
\beq
Q = -2 \log \left(\frac{L_m(s + b)}{L_m(b)}\right) \ .
\eeq
We calculate limits using the $\text{\acro{CL}}_s$~\cite{JUNK1999435,Read_2002} method. We utilize a toy Monte Carlo technique to find the one-sided $p$-value of the observed data in the signal-plus-background and background-only hypotheses, denoted by $p_{s+b} = P(Q_{s+b}\geq Q_{\text{obs}})$ and $p_b=P(Q_b\leq Q_{\text{obs}})$ respectively. We define $\text{\acro{CL}}_s$ to be $\text{\acro{CL}}_s= p_{s+b} / (1-p_b)$.  We set upper limits on Br at 95\% confidence by requiring $\text{\acro{CL}}_s < 0.05$. 

We summarize the limits derived from this method in Fig.~\ref{fig:CMS:limits}.\footnote{We neglect some of the bins with lower significance to make this procedure computationally tractable.}.
Bounds on channels with $\tau$ final states are constrained to roughly the level of Br~$\sim 10^{-2}$, while the $e\mu$ channels are constrained to the level of Br~$\sim 10^{-4}$.  In the $\tau$ cases, most of the events fall into higher background bins, notably the 3-leptons, 1-$\tau_h$ and 1-\acro{OSSF} (0-\acro{OSSF}) of Ref.~\cite{Khachatryan:2016iqn} for Model \ModelII\, (\ModelIV).  As one would expect, the real $\varphi$ models (\ModelIII\, and \ModelIV) are more constrained than their complex $\varphi$ counterparts.  The limits from $Z$-mediated processes are scaled into the Higgs Br parameter space using: 
\beq
\label{eq:z2h}
\text{Br} (H \to \varphi \ell_1 \ell_2 )=  \text{BR} (Z \to \varphi \ell_1 \ell_2 ) \cdot \frac{\Gamma(Z \to \varphi \ell_1 \ell_2 )}{\Gamma(H \to \varphi \ell_1 \ell_2 )} \cdot \frac{\Gamma_h}{\Gamma_Z}
\eeq
where we have used $\Gamma_h=4.07$~MeV~\cite{PDG:Higgs:2018}, $\Gamma_Z=2.49$~GeV~\cite{PDG:Z:2018} as the decay widths of the Higgs and $Z$-boson, respectively. 

%%%%%%%%%%%%%%%%%%%%%
\subsection{A dedicated Higgs search}
\label{sec:search}

We present a targeted method to probe the parameter space of lepton-flavor violating mediators at the \acro{LHC} with  $\sqrt{s} = 13$~TeV and a luminosity of $150$~fb$^{-1}$.  In section~\ref{sec:Multi-lepton:search:limits}, we showed that the limits from $Z$-boson-mediated diagrams are much weaker than those from the Higgs-mediated processes. As such, we ignore them and focus on the Higgs processes in the rest of this section. 
We use $\sigma(pp\to h) = 55.1$ pb~\cite{PDG:Higgs:2018} as the 13~TeV production cross-section of the Higgs.
Since the Higgs is a resonance, we add  a cut on the total invariant mass of the leptons ($M_4$).  This will boost the signal-to-background ratio by eliminating almost all of the background events.

We follow a similar cut procedure as in Section~\ref{sec:Multi-lepton:search:limits}~\cite{Khachatryan:2016iqn}, with the exception that we now relax the $p_T^{\text{sub}}$ cut for the sub-leading light leptons to $7$~GeV instead of $10$~GeV. 
We restrict our study to signatures with four leptons successfully identified at the collider.  In $\tau$ systems, we require one hadronically decaying $\tau_h$ and one leptonic tau with the different flavor than the other two hard leptons. For example, for non-zero $y_{23}$ coupling we require an electron.  We additionally impose a $Z$-window cut on \acro{OSSF} pairs of $\abs{m_Z-m_{\ell^+\ell^-}}>15$~GeV and apply a final cut on the total invariant mass ($M_4$) as defined in Table~\ref{tab:search:region}.

%---------------------------------------------
\begin{table}[t]
  \centering
   \begin{tabular}{@{}r@{\hskip 2em}l@{\hskip 2em}l@{\hskip 2em}l@{}}
   \toprule 
   Model & Required Final States & $M_4$ Cut Range (GeV) \\
   \midrule
   \ModelI    & $e^+\, e^-\, \mu^+\, \mu^-$       &  $120-130$ \\
   \ModelII   & $\tau_h^{\pm}\, e^{\mp}\, \mu^+\, \mu^-$      &  $80 -120$ \\
   \ModelIII  & $e^+\, e^+\, \mu^-\, \mu^-$       &  $120-130$ \\
   \ModelIV   & $\tau_h^-\, e^-\, \mu^+\, \mu^+$  &  $80 -120$ \\
   \bottomrule
   \end{tabular}
  \caption{The four pure off-diagonal, single-coupling models considered in this section.  The opposite-sign final states in model \ModelIII\, and \ModelIV\, are included in our analysis. 
  }
  \label{tab:search:region}
 \end{table}
%---------------------------------------------

We can use the number of events passing all of the cuts to place limits on the branching ratios. However, the signal-to-background ratio can be further boosted by binning the events with respect to the lepton pair invariant masses $(m_{l_1^+ l_2^-}, m_{l_3^+ l_4^-})$ in cases with $\kappa\neq 0$ (see Fig.~\ref{fig:diags}). In model \ModelI, we can record the invariant masses of $e^+\mu^-$, $e^-\mu^+$ pairs. For model \ModelIII\, there is an ambiguity when matching the leptons since there are two ways $e-\mu$ can be paired. We expect the correct combination to have the same invariant masses ($m_{\varphi}$) in an ideal measurement. Therefore, from the two possible combinations ($a$, $b$) we choose the one with the minimum difference in the invariant masses: 
\beq
\begin{split}
\label{eq:SSpair:cond}
a: m_{a_1} &= m_{e_1^+\mu_1^-},\quad m_{a_2} = m_{e_2^+\mu_2^-}, \quad \Delta m_a = |m_{a_2} - m_{a_1}| \\
b: m_{b_1} &= m_{e_2^+\mu_1^-},\quad m_{b_2} = m_{e_1^+\mu_2^-},\quad \Delta m_b = |m_{b_2} - m_{b_1}| \\
& \longrightarrow \qquad \Delta m = \text{Min}(\Delta m_a, \Delta m_b).
\end{split}
\eeq

%---------------------------------------------
\begin{figure}[t]
  \centering
  \includegraphics[width=0.49\textwidth]{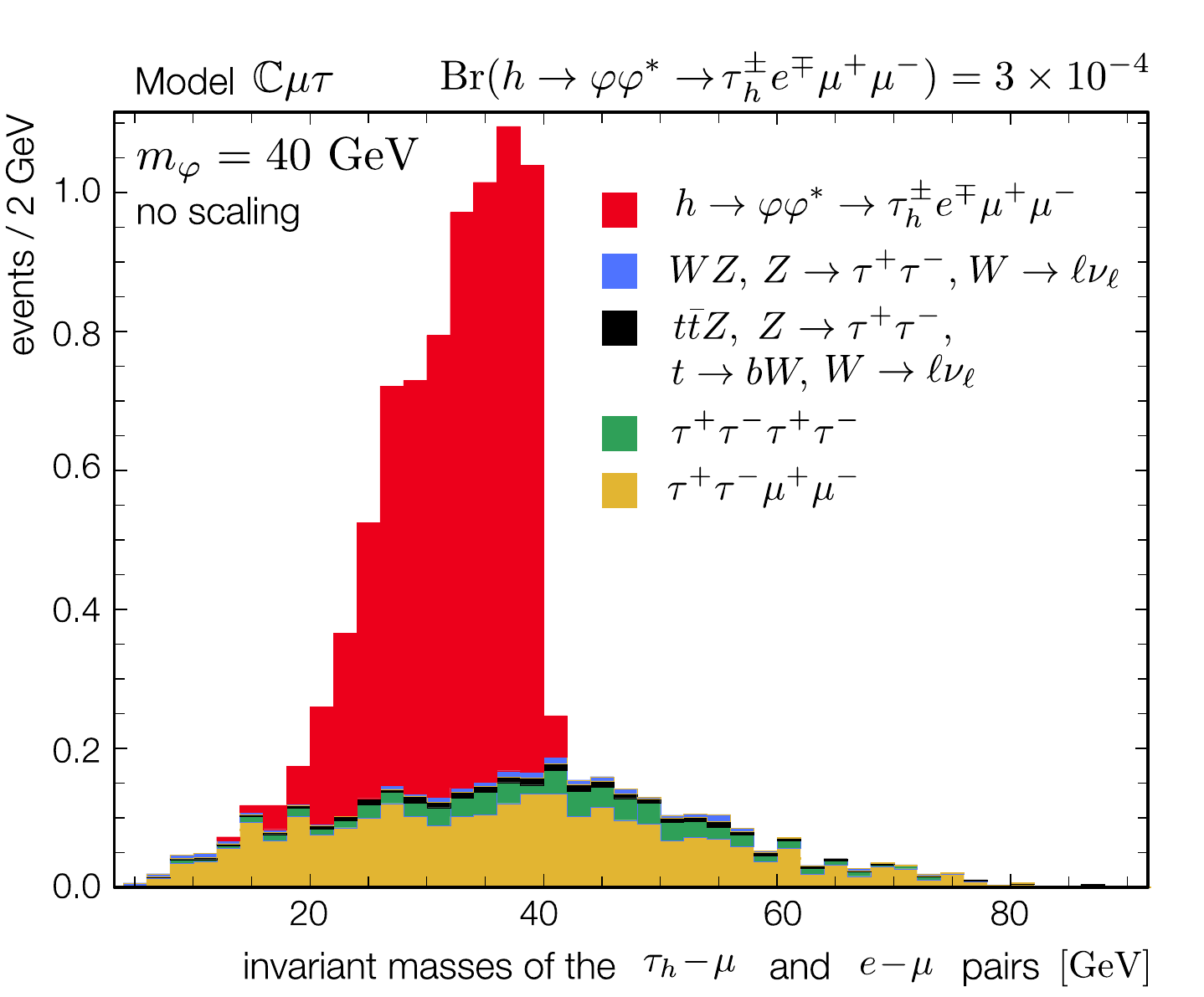}
  \hfill
  \includegraphics[width=0.49\textwidth]{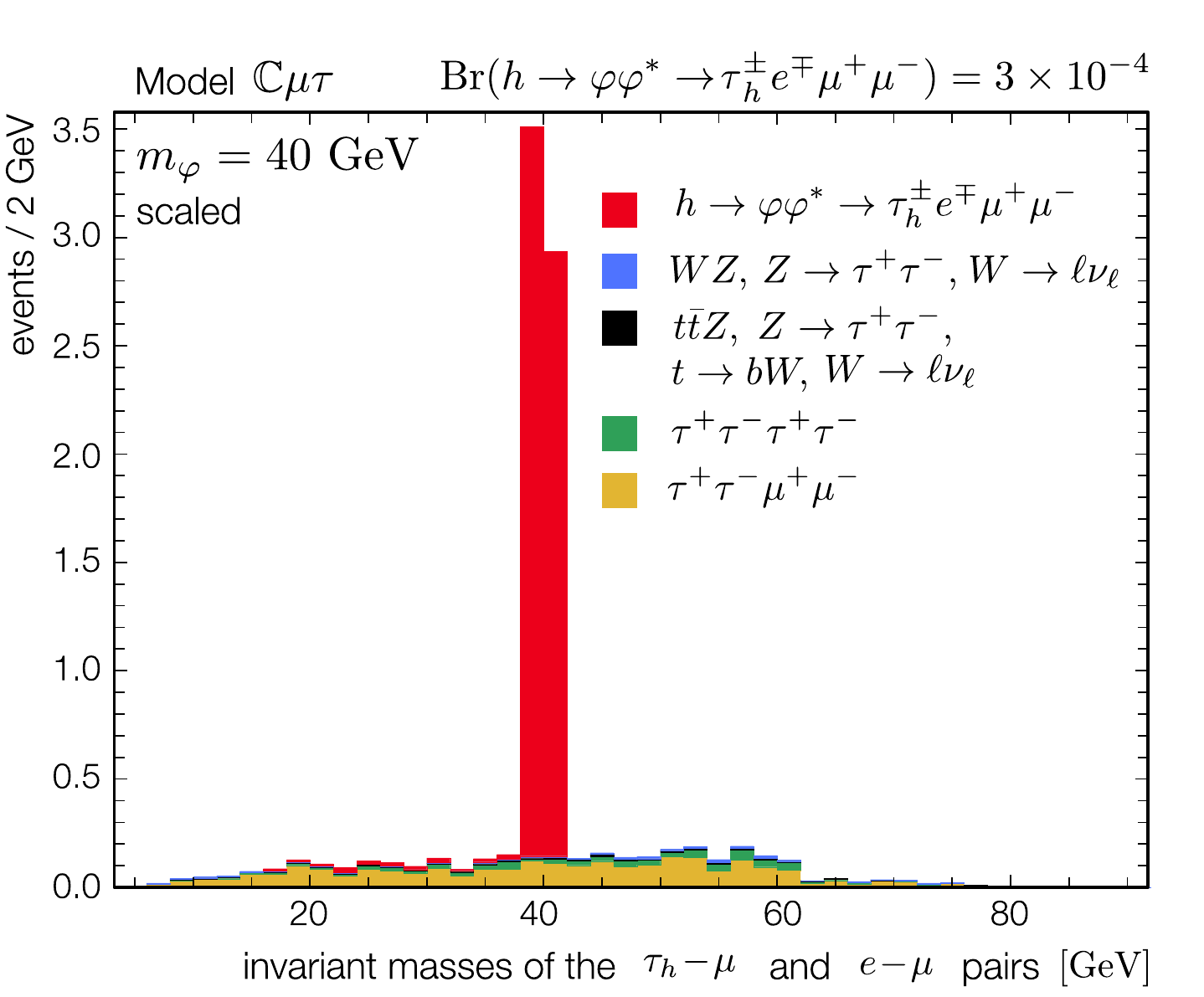}
  \caption{The distribution for invariant masses of $(\tau_h, \mu)$ and $(e, \mu)$ pairs in model \ModelII\, for the case with $\kappa \neq 0$ for both signal (red) and background channels prior to (\textbf{left}) and after (\textbf{right}) the scaling procedure.  In each figure, the mass for both $(\tau_h, \mu)$ and $(e, \mu)$ are shown.  }
  \label{fig:search:invM}
\end{figure}
%---------------------------------------------

In model \ModelII, the invariant masses of the lepton pairs are less than $m_{\varphi}$ due to the missing energy in $\tau$ decays. In order to approximately correct for this missing energy, we perform a momentum scaling of the decay products as described below. Assuming that the $\tau$s are highly boosted, the missing  and visible momenta from the $\tau$ decay ($\vec{p}^{\rm miss}_{\tau}$ and $\vec{p}^{\rm vis}_{\tau}$, respectively) are approximately collinear. Therefore, the true $\tau$ momentum can be written as $p_{\tau} = p^{\rm miss}_{\tau} + p^{\rm vis}_{\tau} = \alpha\, p^{\rm vis}_{\tau}$, where $\alpha$ is a scaling constant that we wish to determine. This visible momentum ($p^{\rm vis}_{\tau}$) is equal to the electron's momentum ($p_e$) in the case of leptonic $\tau$ decays, i.e. $\tau \to e \bar{\nu}_e \nu_{\tau}$, so we implement a scaling of the $\tau_h^{\pm},\, e^{\mp}$ four-momenta:
\begin{align}
\label{eq:mom:scale}
p^{\rm vis}_{\tau_h} &\longrightarrow p_{\tau_h}= \alpha_{\tau_h} p^{\rm vis}_{\tau_h} 
&
\alpha_{\tau_h} &= \left(\frac{m_X}{m_{\tau_h \mu}}\right)^2
\\
p^{\rm vis}_{\tau_\ell} = p_{e} &\longrightarrow p_{\tau_\ell}= \alpha_{e} p_{e} 
&
\alpha_{e} &= \left(\frac{m_X}{m_{e\mu}}\right)^2
\ .
\end{align}
We can determine $m_X$, which can be identified with $m_{\varphi}$, by imposing the total invariant mass constraint,
\beq
\label{eq:mh:scale}
\text{Inv}(e', \mu, \tau_h', \mu) \approx m_h = 125 \text{ GeV} \ ,
\eeq
where the four-momenta for $e'$ and $\tau_h'$ are scaled as in \eqref{eq:mom:scale}. After solving \eqref{eq:mh:scale} for $m$ and plugging it back into \eqref{eq:mom:scale}, we derive the scaled momenta and use them for binning the events.
This scaling is justified because in the relativistic limit for the $e-\mu$ pair, we have $m_{e\mu} \approx \sqrt{2\, p_e\cdot p_\mu}$. A scaling of $p_e$ by $\alpha_e$ results in a scaling $\sqrt{\alpha_e}$ in $m_{e\mu}$. As for the $\tau_h-\mu$ pair, the ratio of the scaled invariant mass $(p_{\tau_h}+p_\mu)^2 = m'^2_{\tau_h\mu}$ to the unscaled invariant mass $(p_{\tau_h}^{\text{vis}}+p_\mu)^2 = m_{\tau_h\mu}^2$ is:
\beq
\label{eq:pscale:appr}
\left(\frac{m'_{\tau_h\mu}}{m_{\tau_h\mu}} \right)^2 \approx \frac{\alpha_{\tau_h}^2 m_{\tau_h}^2 + 2\alpha_{\tau_h} p_{\tau_h}\cdot p_{\mu}}{ m_{\tau_h\mu}^2} \approx \alpha_{\tau_h} + \left(\frac{m_{\tau_h}}{m_{\tau_h\mu}}\right)^2 \alpha_{\tau_h}(\alpha_{\tau_h} - 1) \ .
\eeq
If we take $m_{\tau_h \mu}\approx 10$~GeV and also assume that the scaling factor is close to one, the sub-leading terms in~\eqref{eq:pscale:appr} are  much less than $10^{-3}$.  The result of this scaling for a sample case ($m_{\varphi}= 40$~GeV) is shown in Fig.~\ref{fig:search:invM}.  This scaling method is exceptional at sharpening the signal without inducing any spurious focusing of the background distribution. 

In model \ModelIV, due to the missing energies we can no longer resolve the ambiguity in matching the leptons by using the invariant masses, so we simply use the number of events passing all the cuts to set limits in this case. 
However, the scaling procedure outlined above can still be performed to find the location of a bump, i.e.\ $m_{\varphi}$, in the event of an excess. The scaling should be done for both possible combinations of $(\tau_h-\mu$, $e-\mu)$ pairs. The solutions to~\eqref{eq:mh:scale}, $\alpha^{1, 2}_{\mu,\tau}$, must be checked. If either $\alpha_{\tau_h}$ or $\alpha_{e}$ in a combination is less than one, the combination is incorrect and is discarded. The cases with more than one correct scaling combination are ambiguous, but the $m_\varphi$ feature should still emerge clearly. Alternatively, one could use only the unambiguous set of scaled invariant masses to determine the mediator mass.

The scaling method works well for $h\to \varphi\varphi^*$, but it is not able to enhance sensitivity to the $h\to \varphi \ell_i^+ \ell_j^-$ signals as there is only a single resonance.  However, as can be observed on the left side of Fig.~\ref{fig:search:invM}, the distribution does create a kinematic endpoint in the invariant mass which could be used to ascertain the internal resonance mass.

%---------------------------------------------
\begin{table}[t]
  \centering
   \begin{tabular}{@{}r@{\hskip 2em}l@{\hskip 2em}l@{\hskip 2em}l@{}}
   \toprule 
   Model & Process & Signal Region (GeV) \\
   \midrule
   \ModelI    & $h\to \varphi \varphi^*$       &  $m_{e^+\mu^-}\, \text{ and }\, m_{e^-\mu^+} \in [m_{\varphi} - 3, m_{\varphi} + 3]$ \\
   \ModelI    & $h\to \varphi e^+ \mu^-$       &  $m_{e^+\mu^-}\, \text{ or }\, m_{e^-\mu^+} \in [m_{\varphi} - 3, m_{\varphi} + 3]$ \\
   \ModelII    & $h\to \varphi \varphi^*$       &  $m'_{\tau_h^{\pm}\mu^{\mp}}\, \text{ and }\, m'_{e^{\mp}\mu^{\pm}} \in [m_{\varphi} - 5, m_{\varphi} + 5]$ \\
   \ModelII    & $h\to \varphi \tau^+ \mu^-$       &  All values of \, $m_{\tau_h^{\pm}\mu^{\mp}}\, \text{ and }\, m_{e^{\mp}\mu^{\pm}}$ \\
   \ModelIII    & $h\to \varphi \varphi^*$       &  $m_{e^+\mu^-}\, \text{ and }\, m_{e^+\mu^-} \in [m_{\varphi} - 3, m_{\varphi} + 3]$ \\
   \ModelIII    & $h\to \varphi e^+ \mu^-$       &  All values of \,$m_{e^+\mu^-}\, \text{ and }\, m_{e^+\mu^-}$ \\
   \ModelIV    & Both       &  All values of \,$m_{\mu^+\tau_h^-}\, \text{ and }\, m_{\mu^+e^-}$ \\
   \bottomrule
   \end{tabular}
  \caption{Definition of signal bins used to place limits on the branching ratios in each case. For the \acro{OSSF} events in model \ModelIII\, and \ModelIV, we use the corresponding definition from model \ModelI\, and \ModelII.
  }
  \label{tab:search:megabin}
 \end{table}
%---------------------------------------------

%---------------------------------------------
\begin{figure}[t!]
  \centering
  \includegraphics[width=0.49\textwidth]{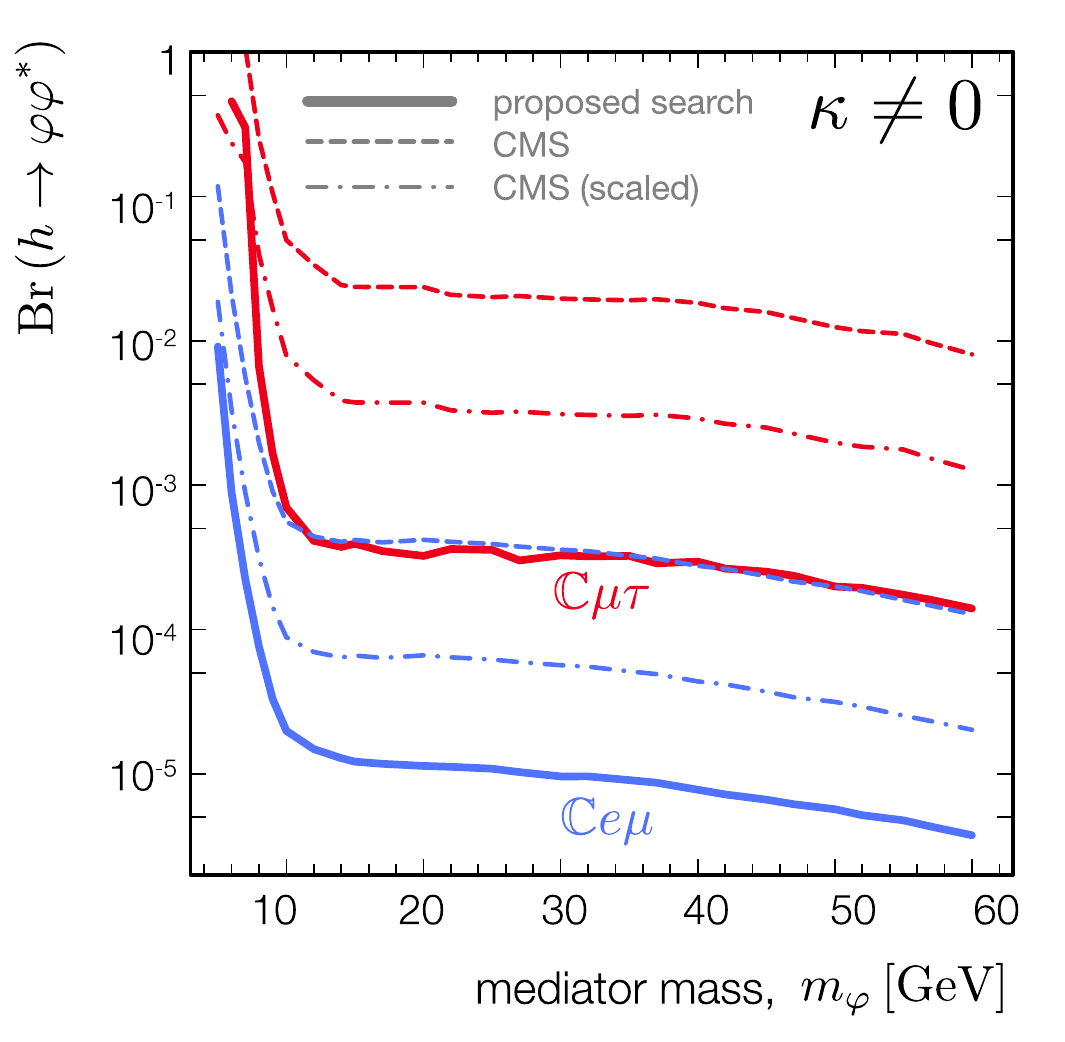}
  \hfill
  \includegraphics[width=0.49\textwidth]{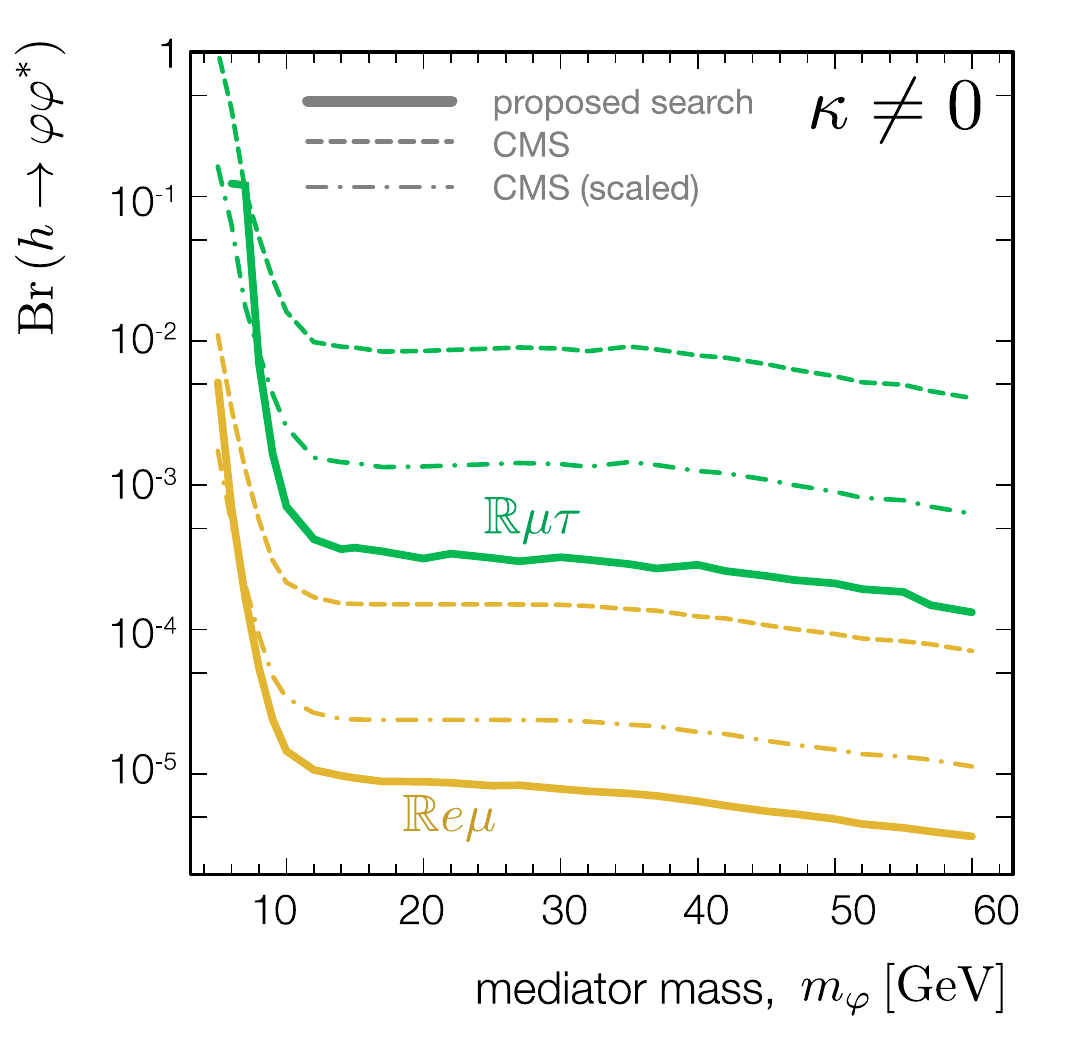}
  \hfill
  \includegraphics[width=0.49\textwidth]{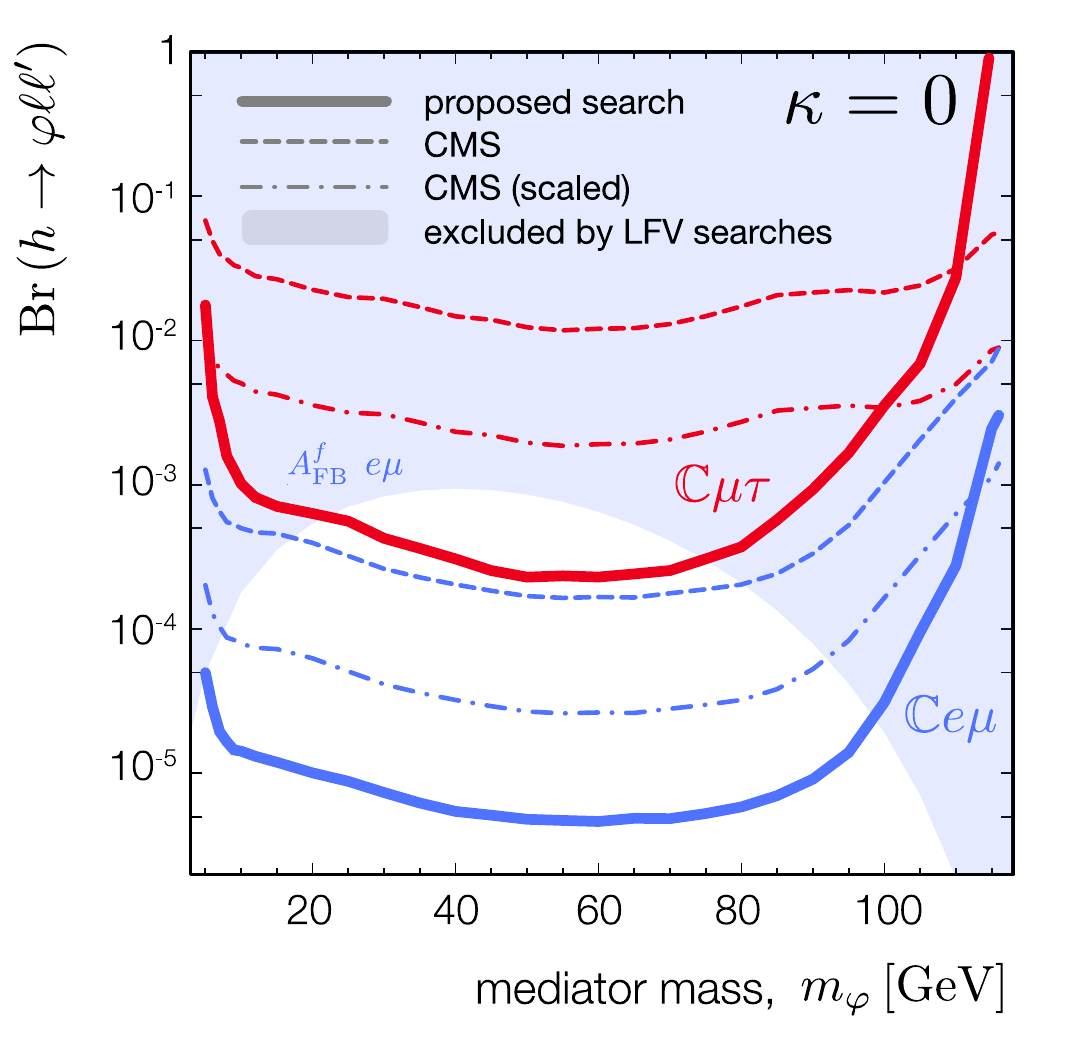}
  \hfill
  \includegraphics[width=0.49\textwidth]{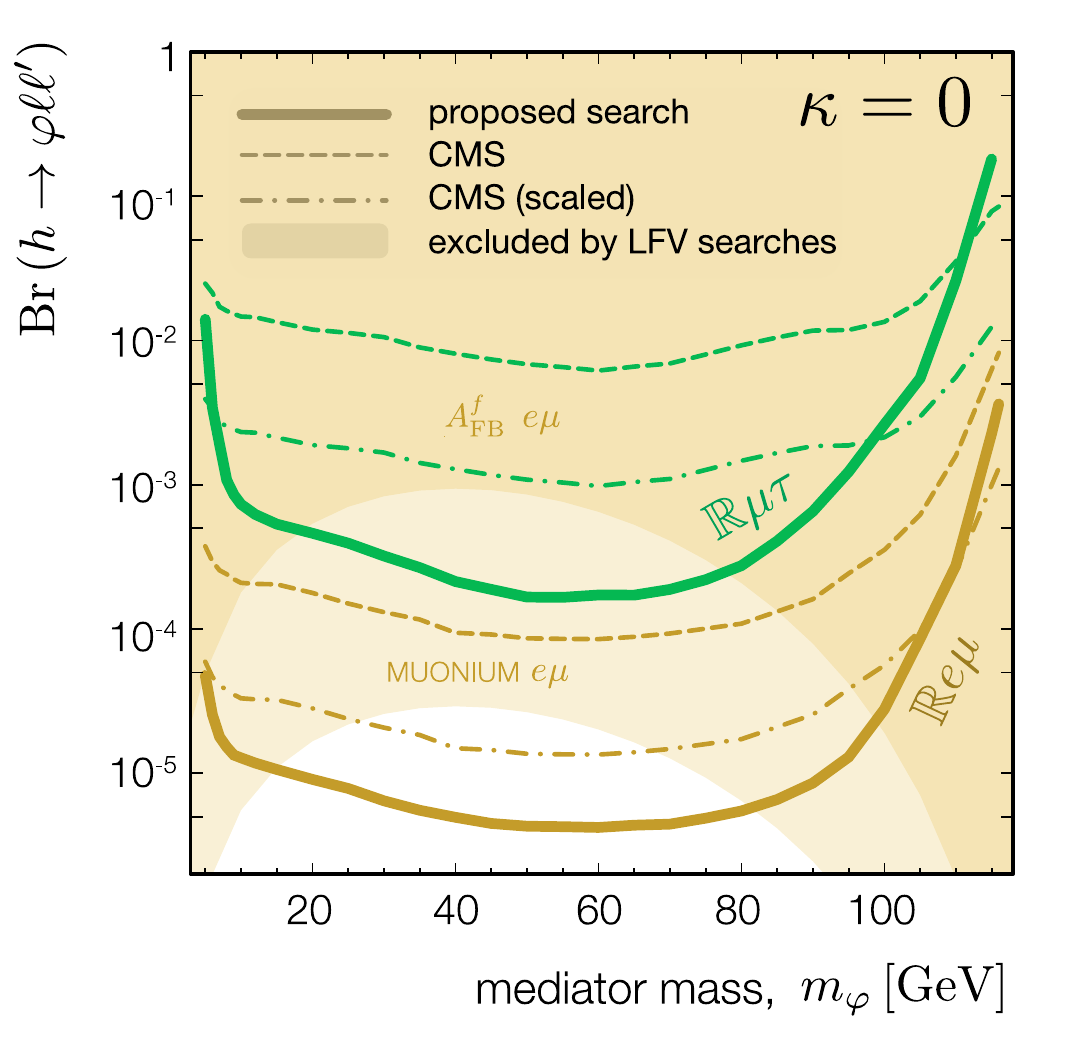}
  \caption{Limits on the branching ratio at 95\% \acro{CL} for the models listed in Table~\ref{tab:models} in the case with a non-zero quartic coupling $\kappa\neq 0$ (\textbf{top}), and a zero quartic coupling $\kappa = 0$ (\textbf{bottom}). The dashed curves are \acro{CMS}~results at 8~TeV~\cite{Khachatryan:2016iqn}. The dot-dashed curves are the same results scaled using~\eqref{eq:search:cms8to13}. The thick curves correspond to our search proposal at $\sqrt{s} = 13$~TeV, and a luminosity of 150~fb$^{-1}$.  The bottom two figures show the excluded region from the most relevant lepton-flavor violating searches in the $e\mu$ cases.  The $\mu\tau$ cases are unconstrained.}
  \label{fig:search:c1:limits}
\end{figure}
%---------------------------------------------

We consider three background channels for model \ModelI\,and \ModelIII: (a) $p p \to e^+ e^- \mu^+ \mu^-$, (b) $t\bar{t}Z$,  (c) $W^+W^-Z$, where $t\to Wb$, and $W/Z$ bosons decay leptonically. The dominant background in model \ModelI\, is from $p p \to e^+ e^- \mu^+ \mu^-$, while the 2-\acro{SSSF} background in model \ModelIII\, is negligible. For model \ModelII\, and \ModelIV, we analyzed the background from (a) $p p \to \mu^+ \mu^- \tau^+ \tau^-$, (b) $\tau^+ \tau^- \tau^+ \tau^-$, (c) $t\bar{t}Z$,  (d) $W^+W^-Z$ processes. The background originating from jets misidentified as $\tau_h$ (``fake taus") in three main channels (a) $t\bar{t}W$, (b) $WZ$, (c) $t\bar{t}Z$ are also included. In order to estimate this fake background, we first generate events with $n$ jets $ +\, 3$ light leptons final states with {\tt FastJet-3}~\cite{Cacciari:2011ma} using anti-kT  algorithm~\cite{Cacciari:2008gp} with a distance parameter $R = 0.4$. We then form a set of $n$ events corresponding to each jet, which are weighted using the jet $\to \tau_h$ misidentification rate~\cite{Sirunyan:2018pgf}, which is typically of $\order{1\%}$. Finally, we normalize the background events in all cases by 
\beq
\label{eq:bg:weight}
w =  \frac{\mathcal{L} \cdot \sigma}{ N} \cdot K
\eeq
where $\sigma$ is the production cross-section for each specific background channel at $\sqrt{s}=13$ TeV and we take $\mathcal{L} = 150$ fb$^{-1}$. We also assume the K-factor $\sim 1.7$~\cite{Grazzini:2018owa} to correct for the \acro{NNLO} effects. The distribution of invariant masses of $\tau_h-\mu$ and $e-\mu$ pairs (for both signal and background) with an example choice of parameters in model \ModelII\, are shown in Fig.~\ref{fig:search:invM}. It can be seen that the cut procedure eliminates most of the background events in this case, and similarly in other models. 

We follow the same statistical methods to set limits on the branching ratios, with a different binning procedure. We used one bin for models with \acro{OSSF} signatures, and two bins for model \ModelIII\, and \ModelIV, i.e. for the \acro{OSSF} and \acro{SSSF} contributions. The definition of these signal bins is presented in Table~\ref{tab:search:megabin}. 

In Fig.~\ref{fig:search:c1:limits}, we summarize the projected limits from our search proposal, reproduce the \acro{CMS} search at $\sqrt{s}=8$ TeV~\cite{Khachatryan:2016iqn} (Section~\ref{sec:Multi-lepton:search:limits}), and add to these the same \acro{CMS} limits na\"ively projected to $\sqrt{s}=13$ TeV and $150$ fb$^{-1}$. These projected \acro{CMS} bounds are estimated by
\beq
\label{eq:search:cms8to13}
\text{BR}_{13} = \text{BR}_{8} \cdot \sqrt{\frac{19.5 \text{ fb}^{-1}}{150 \text{ fb}^{-1}}} \cdot \frac{\sigma_{8}(pp \to h)}{\sigma_{13}(pp \to h)}
\eeq
where $\sigma_{13}(pp\to h) = 55.1$ pb~\cite{PDG:Higgs:2018}, and $\sigma_{8}( p p \to h) = 24.2$ pb~\cite{PDG:Higgs:2018}.

%%%%%%%%%%%%%%%%%%%%%
\subsection{Long-lived LFV scalars}
\label{sec:LLP}

Throughout this work, we have assumed that $\varphi$ decays promptly in the detector.  In principle, it could be long-lived and yield a displaced signature in the detector.  The lifetime of $\varphi$ is 
\beq
\Gamma(\varphi \to \ell_i^+\ell_j^-) =\frac{m_\varphi}{16\pi}\left[  (\abs{y_{ij}}^2+\abs{y_{ij}'}^2) (1- r_i^2 -r_j^2)-4r_ir_j \text{Re}(y_{ij}^*y_{ij}') \right] \lambda^{1/2}(1,r_i^2,r_j^2),
\eeq
where $r_i = m_{\ell_i}/m_\varphi$ and $\lambda(a,b,c) = a^2+b^2+c^2-2(ab +bc+ac)$ is the phase space factor.  Assuming a single, off-diagonal coupling $y_{ij}\neq0$, the characteristic displacement scale is
\beq
c\tau_\varphi \approx 500 \text{ $\mu$m} \lp\frac{10 {\text{ GeV}}}{m_\varphi}\rp \lp\frac{10^{-6}}{\abs{y_{ij}}}\rp^{2} \frac{\lambda^{-1/2}(1,r_i^2,r_j^2)}{1- r_i^2 -r_j^2}.
\eeq

Targeted searches are typically much more sensitive to long-lived particles than their prompt counterparts. This is especially true with leptons, as impact parameter criteria designed to remove cosmic muons, material interactions, and other rare backgrounds run the danger of removing these non-prompt leptons from the signal entirely~\cite{Evans:2016zau,Alimena:2019zri}.  Many \acro{LHC} searches exist for long-lived particles.  Of those in the lepton channels, the $p_T$ requirements at the trigger level can be quite harsh, removing sensitivity to Higgs decays, e.g.~\cite{Khachatryan:2014mea,Aad:2015rba,Aad:2019tcc}, others are focused on displaced $ee$ or $\mu\mu$ signatures, e.g.~\cite{CMS:2014hka,Aad:2014yea}, and others target fairly specific signatures~\cite{Aad:2019kiz}.  Recasting these studies is beyond the scope of this work, but we note that scalars with an~\acro{LFV} decay in these models can easily be displaced and are not optimally constrained by existing searches.

%%%%%%%%%%%%%%%%%%%%%%%%%%%%%%%%%
\section{Conclusion}
\label{sec:conclusions}

Exotic Higgs decays are one of the most promising places to uncover new physics in the near future.  In order to ensure no signals gets overlooked, it is essential to have a comprehensive program.  In this work, we explored a simple model containing a new scalar lighter in mass than the Higgs that decays into two standard model charged leptons of different flavor.  Although there is substantial motivation to consider new physics within the lepton sector from many extant anomalies, this specific signature had been overlooked thus far at the \acro{LHC}.

At the weak scale, the \acro{LFV} decay of the new scalar originates from dimension-5 operators involving the Higgs, which also facilitates a decay path for the Higgs $h \to \varphi\ell\ell'$.  The operator can originate from a variety of simple ultraviolet completions.   The addition of a dimension-4 coupling allows the direct decay $h \to \varphi\varphi^*$.  In the case of a real scalar, this can result in an exotic signature with two same-sign same-flavor pairs of leptons, e.g.~$e^+e^+\mu^-\mu^-$.

New \acro{LFV} couplings can impact many precision and flavor observables.  We show that constraints from these prove very mild if the flavorful couplings to leptons are completely off-diagonal, with the most stringent being forward-backward asymmetries for couplings involving electrons, and muonium oscillations for a real scalar with $y_{12}\neq 0$.  In the case where the alignment is not exact, additional observables can be constraining, notably lepton radiative decays and decays into three charged leptons.  Prompt scalar decays can very easily be accommodated within these constraints.

An existing low missing energy and low $S_T$ multi-lepton search can be used to place constraints on this model, but a dedicated search that capitalizes on the distinctive kinematics can perform much better.  In such a search, requiring a four-lepton invariant mass consistent with a Higgs parent allows for a rejection of most of the small backgrounds.  Even in the cases with taus, information can be gleaned about the mass of the new scalar through application of kinematic end points or by enforcing a Higgs mass constraint.  While prompt signatures were the focus of this work, the new scalars could instead be displaced.  Such a striking displaced signal with relatively soft leptons could potentially benefit from a dedicated search.

%%%%%%%%%%%%%%%%%%%%%%%%%%%%%%%%%
\section*{Acknowledgements}

We thank
J.~Brod, I.~Galon, K.C.~Kong, D.~McKeen, J.~Shelton, and J.~Zupan 
for useful comments and discussions.
We also thank Y.C.~Ding for his assistance with technology.
\textsc{j.a.e.} and \textsc{p.t.} thank the Aspen Center for Physics (\acro{NSF} \acro{PHY}-{\small{1066293}) for its hospitality when this work was initiated. \textsc{p.t.} thanks the Kavli Institute for Theoretical Physics (\acro{NSF} \acro{PHY}-{\small{1748958}}) for its hospitality while part of this manuscript was being completed.
\acro{J.A.E.}~acknowledges support by \acro{DOE}~grant \acro{DE-SC}{\small{0011784}}. \acro{P.T.}~is supported by the \acro{DOE}~grant \acro{de-sc}{\small{0008541}}. \textsc{m.z.}~is supported by the \acro{CAS} \acro{PIFI} grant {\small{\#2019PM0110}}.

\appendix

%%%%%%%%%%%%%%%%%%%%%%%%%%%%%%%%%%%%
\section{Renormalizable ultraviolet completions}
\label{app:UV:completion}

The dimension-5 weak-scale effective operator \eqref{eq:dim:5} can emerge from renormalizable theories whose additional degrees of freedom have been integrated out. We present three simple example models that do not introduce additional low-energy states that would influence the phenomenology.

%%%%%%%%%%%%%%%%%%%%%%%%%%%%%%%%%%%%
\subsection{Vector-like leptons}

One of the simplest extensions is to introduce a vector-like pair of new leptons, $\xi_{L,R}$,
 \begin{align}
  \mathcal{L}^V_{\xi}
  &\supset
  M \bar \xi_L\xi_R + \left[\lambda_i\, \bar L_i \cdot H \xi_R +  \lambda_i'\, \bar \xi_L E_i  \varphi  + \hc \right].
\end{align}
We have chosen the electroweak quantum numbers of $\xi_{L,R}$ such that $\xi_R$ and $E$ have the same charges. Because the new fermions form a vector-like pair, the Dirac mass $M$ may be naturally large. The $y_{ij}$ in \eqref{eq:fi:sm} are generated by 
a tree-level diagram with virtual heavy leptons. In turn, the low-energy effective interactions in \eqref{eq:fi:sm} appear upon inserting the Higgs vacuum expectation value. For example,
\begin{align}
  \vcenter{
    \hbox{\includegraphics[height=.15\textwidth]{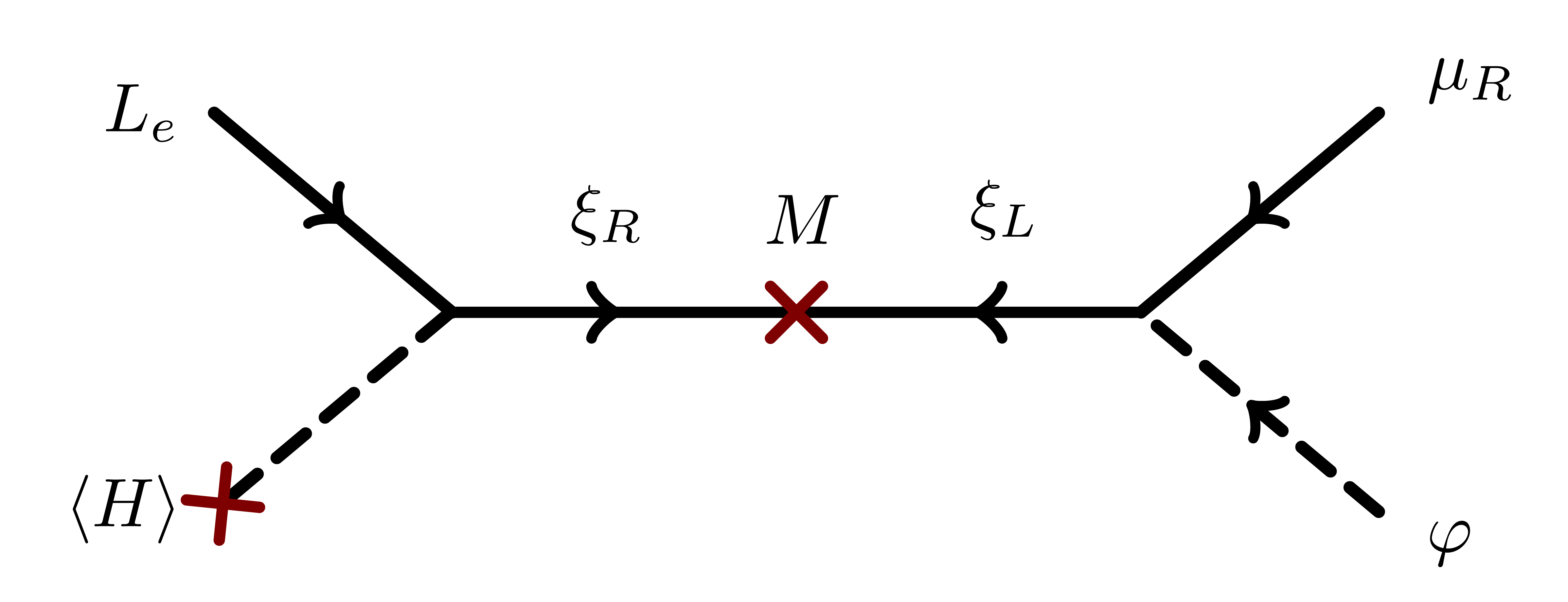}}
    }
  & \Longrightarrow
  \vcenter{
    \hbox{\includegraphics[height=.15\textwidth]{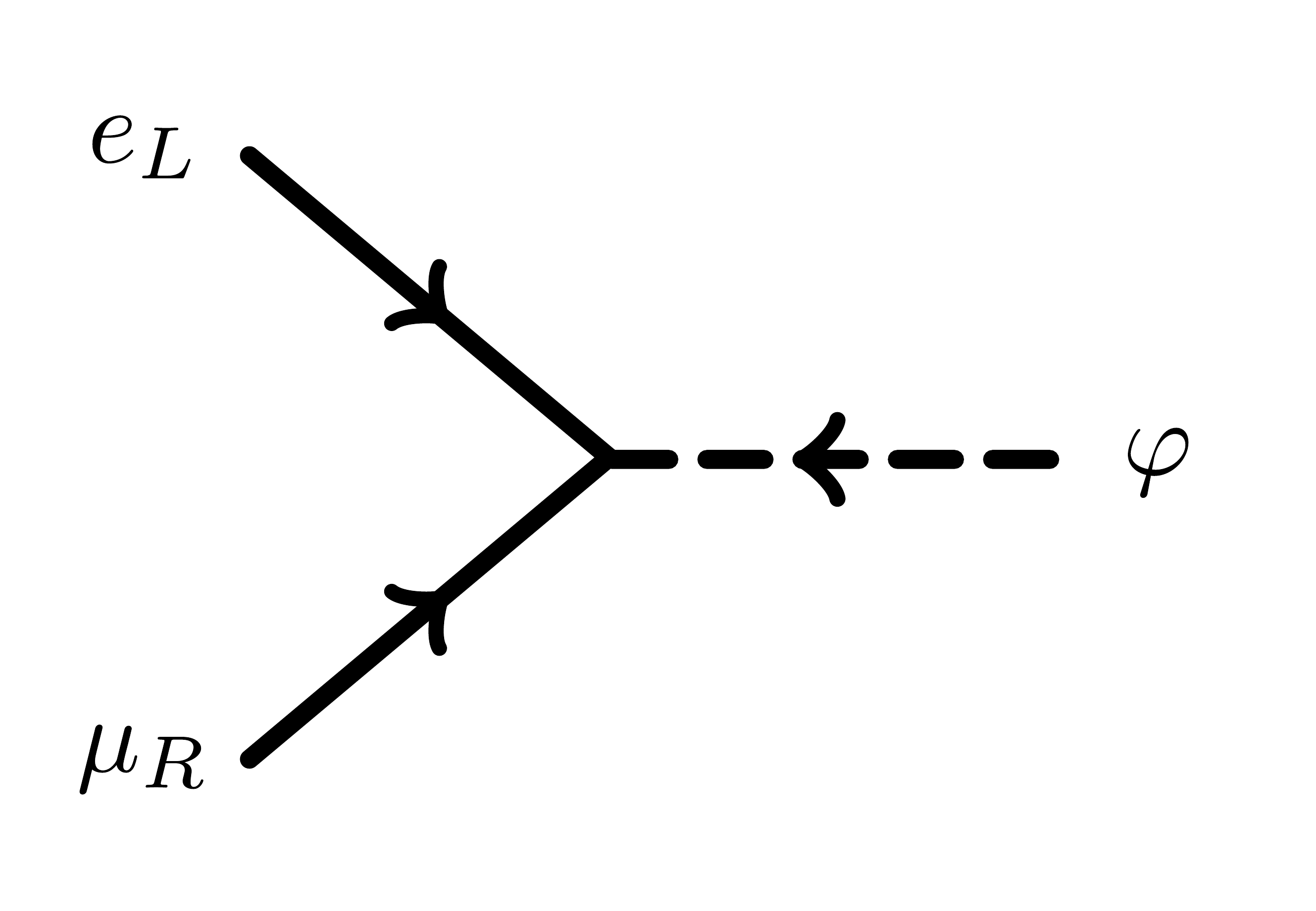}}
    }
    \ ,
\end{align}
where fermion arrows indicate helicity and the $\varphi$ arrow indicates $L_e-L_\mu$ charge.
The effective scale in \eqref{eq:dim:5} is then identified with the mass of the heavy fermion,  $\Lambda \sim M$. Ref.~\cite{Batell:2017kty} recently explored the loop-level implications of this class of ultraviolet completion.  In this model, the off-diagonal nature is not manifestly enforced.  Essentially, one needs to introduce a $\lambda$ and $\lambda'$ that are differently flavor directed and well aligned with the mass eigenbasis in flavor space, e.g.~$\lambda\sim(1,0,0)$ and $\lambda'\sim(0,1,0)$.

%%%%%%%%%%%%%%%%%%%%%%%%%%%%%%%%%%%%
\subsection{Froggatt--Nielsen}

The Froggatt--Nielsen mechanism generates the Standard Model fermion mass hierarchy by integrating out heavy degrees of freedom that break Abelian flavor symmetries~\cite{Froggatt:1978nt}. We may apply this framework with a single $U(1)_X$ flavor symmetry and heavy scalar, $S$,
\begin{align}
 \mathcal L_{\text{Lep.}}^{\text{FN}}
 &\supset
 \alpha_{ij}\,  \bar L_i \cdot H E_j \left(\frac{S}{M}\right)^{|n_{ij}^Y|}.
 \label{eq:fn:lepton}
\end{align}
The scalar field $S$ has $U(1)_X$ flavor charge $[S]_X = -1$ so that the power $n_{ij}^Y = [\bar L_i]_X + [H]_X+ [E_j]_X$. We assume $\alpha \sim \mathcal{O}(1)$, and $M$ is the scale at which the heavy degrees of freedom were integrated out. Assume that $U(1)_X$ is broken at roughly the scale $M$ by $\langle S \rangle \approx 0.2 M$. This generates the Standard Model Yukawa hierarchy for the charged leptons. The Lagrangian in \eqref{eq:dim:5} can be generated similarly, as
\begin{equation}
 \mathcal L_{\varphi\text{-lep.}}^{\text{FN}} \supset \beta_{ij}\,  \bar L_i \cdot H E_j \left(\frac{S}{M}\right)^{|n_{ij}^g|} \frac{\varphi}{\Lambda} + \beta_{ij}'\,  \bar L_i \cdot H E_j \left(\frac{S}{M}\right)^{|n_{ij}^{g'}|} \frac{\varphi^*}{\Lambda},
 \label{eq:fn:med}
\end{equation}
where $n_{ij}^g = [\bar L_i]_X + [H]_X+ [E_j]_X + [\varphi]_X $. For a review of the Froggatt--Nielsen mechanism, and a more phenomenologically realistic example, see Ref.~\cite{Galon:2016bka}, where a product of two $U(1)$ flavor symmetries is implemented.

%%%%%%%%%%%%%%%%%%%%%%%%%%%%%%%%%%%%
\subsection{R-parity violating supersymmetry} 

Finally, one may directly generate \eqref{eq:fi:sm} without the intermediate step \eqref{eq:dim:5} if one identifies the mediator $\varphi$ as a sneutrino in $R$-parity violating supersymmetry~\cite{Halprin:1993zv,Mohapatra:1991ij,Dreiner:1998wm}:
\begin{align}
W_{\slashed{R}_p} &\supset \frac{1}{2}  \sum_{i,j,k}\lambda_{ijk}\, \mathbb{L}_i \cdot \mathbb{L}_j \cdot\bar{\mathbb{E}}_k \ ,
\end{align}
where $i,j,k$ are generation indices, $\mathbb{L}$ is the lepton SU(2)$_L$ doublet superfield, and $\bar{\mathbb{E}}$ is the electron singlet superfield.
Electroweak and flavor symmetries require this term to be anti-symmetric in $\{i,j\}$, i.e. $i\neq j$. If we extract the Yukawa couplings from this term we get
 \begin{align}
  \mathcal{L}_{LL\bar E}
  &= \lambda_{ijk}\,\left[ \tilde{\nu}_L^i\, \bar{e}_R^k e_L^j + \tilde{e}_L^j\, \bar{e}_R^k \nu_L^i  + \left(\tilde{e}_R^k\right)^*\, \left(\bar{\nu}_L^i\right)^c e_L^j - (i\leftrightarrow j)\right] + \hc,
\end{align}
 in which we can identify the mediator $\varphi$ with the sneutrino $\tilde{\nu}_L$.  In this case, the interactions with the $\varphi$ and the Higgs in \eqref{eq:scalar} are generated through $D$-terms and soft terms~\cite[Eq.~(2.5)]{Grossman:1998py}. 
 
 However, this framework suffers from two major problems.  First, it is challenging to decouple the slepton portion of the doublet from the sneutrino.  A light charged slepton is fairly difficult to conceal from a variety of searches.  Second, the sneutrino itself can be amply produced through a $Z$ boson, and, if $m_{\tilde\nu} < m_Z /2$ would appreciably correct the $Z$ width.  Fortunately, a fairly simple resolution would be to introduce a right-handed neutrino superfield $\mathbb N$.  A small $A$-term $H \tilde N \tilde L$ would generate a slight right-handed left-handed sneutrino mixing.  This light right-handed sneutrino is a viable $\varphi$ candidate.  While an interesting possibility, further exploration of this model is well beyond the scope of this work.

%%%%%%%%%%%%%%%%%%%%%%%%%%%%%%%%%%%%%%%%
\section{Review of chiral structure}
\label{app:chiral}

For clarity, we will briefly review the chiral structure of the Yukawa interaction that plays an important role in our model.
  For more details than presented here, see, Martin's 2011 \acro{TASI} lectures~\cite{Martin:2012us} or the comprehensive version with Dreiner and Haber~\cite{Dreiner:2008tw}.   A Dirac fermion $\Psi$ is a mixture of left-handed ($\psi$) and right-handed ($\bar\chi$) Weyl fermions with the same conserved charges, which can be represented in the Dirac basis for the $\gamma$-matrices as
\begin{align}
  \Psi &= 
  \begin{pmatrix}
    \psi \\
    \bar\chi  
  \end{pmatrix}
  &
  \bar\Psi \equiv \Psi^\dag \gamma^0 &= 
  \begin{pmatrix}
    \chi & \bar\psi
  \end{pmatrix} \ . 
\end{align}
In this notation, barred (unbarred) Weyl spinors are understood to be right-(left-)handed. Complex conjugation converts a left-chiral fermion into a right-chiral anti-fermion, so one may consider $\bar\chi = \chi^\dag$ where $\chi$ is the left-handed \acro{CP} conjugate of $\bar\chi$. 
Note that left- and right-chiral spinors have different indices\footnote{In Van der Waerden notation these are written  $\psi_\alpha$ and $\bar\chi^{\dot{\alpha}}$. For our purposes it is sufficient to leave these indices implicit. It is sufficient to know that contractions $\chi\psi = \psi\chi$ and $\bar\chi\bar\psi = \bar\psi\bar\chi$ are allowed, but $\chi\bar\psi$ and $\bar\chi\psi$ are not.} that cannot be contracted with one another as they are in different induced representations of the Poincar\`e group. We explicitly decompose a Dirac electron field, $e$, into its Weyl components $e_L$ and $e_R$:
\begin{align}
    e &= 
    \begin{pmatrix}
      e_L \\
      \bar e_R  
    \end{pmatrix}
    &
    \bar e &=
    \begin{pmatrix}
      e_R & \bar e_L
    \end{pmatrix} \ .
\end{align}
The subscripts $L,R$ differentiate two fundamentally different fields in the Standard Model.
\begin{itemize}
\item $e_L$ is a left-handed electron with charge $Q=-1$ that is part of an electroweak doublet.
\item $\bar e_L$ is its conjugate, a right-handed positron ($Q=+1$) that is part of an electroweak doublet.
\item $\bar e_R$ is a right-handed electron with charge $Q=-1$ that is an electroweak singlet.
\item $e_R$ is its conjugate, a left-handed positron ($Q=+1$) that is an electroweak singlet.
\end{itemize}
The Dirac muon field $\mu$ can be analogously decomposed into Weyl $\mu_{L,R}$ and $\bar\mu_{L,R}$ fields.
With respect to these Weyl fermions, the interactions in \eqref{eq:fi:sm} are
\begin{align}
 y_{12}\bar e P_L \mu \varphi &= y_{12} e_R \mu_L \varphi
 &
 y'_{12} \bar{e} P_R \mu \varphi &= y'_{12} \bar e_L\bar \mu_R \varphi 
 \label{eq:Weyl:verts}
 \\
 \left(y_{12}\bar e P_L \mu \varphi\right)^* &= y_{12}^* \bar e_R \bar \mu_L \varphi^*
 &
 \left(y'_{12} \bar{e} P_R \mu \varphi\right)^* &= \left(y'_{12}\right)^*  e_L \mu_R \varphi^* \ .
 \label{eq:Weyl:verts:star}
\end{align}
These are four distinct interactions, though the reality of the Lagrangian connects the terms in \eqref{eq:Weyl:verts} to those in \eqref{eq:Weyl:verts:star}. These interactions are shown diagrammatically in Figure~\ref{fig:vertices}, where arrows correspond to fermion helicity. 
To aid in translation between the four-component and two-component notation, we write \eqref{eq:fi:sm} in terms of the Weyl fields:
\begin{align}
\mathcal L \supset
\left(
  y_{ij} \ell_{Ri} \ell_{Lj} \varphi + y^*_{ij} \bar\ell_{Ri}\bar\ell_{Lj}\varphi^*
\right)
+
\left(
  y'_{ij} \bar\ell_{Li} \bar\ell_{Rj} \varphi + y'^*_{ij} \ell_{Li}\bar\ell_{Rj} \varphi^*
\right)
  \ .
\end{align}

\begin{figure}
  \centering
  \includegraphics[width=.22\textwidth]{figures/vert_elf.pdf}
  \;
  \includegraphics[width=.22\textwidth]{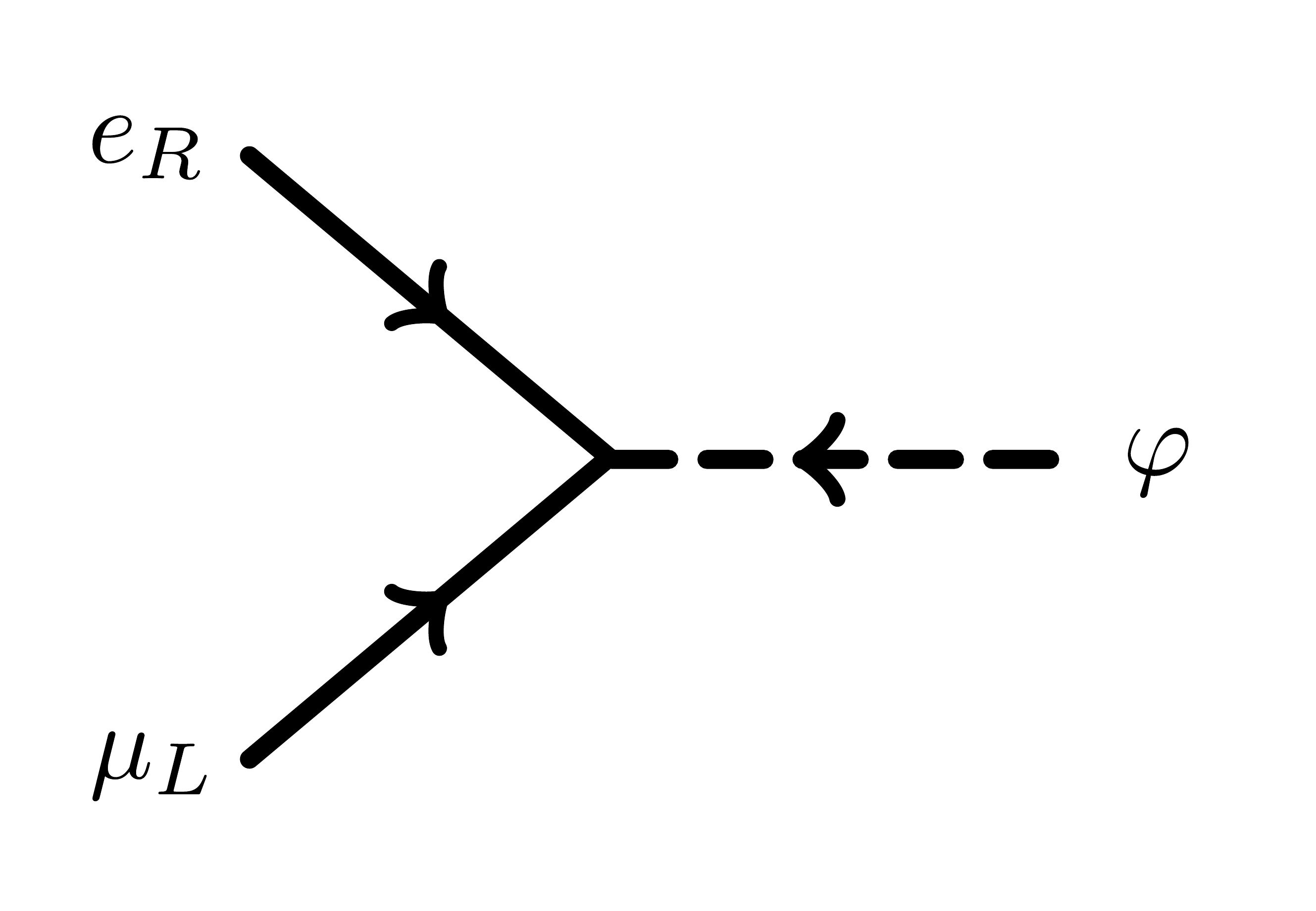}
  \;
  \includegraphics[width=.22\textwidth]{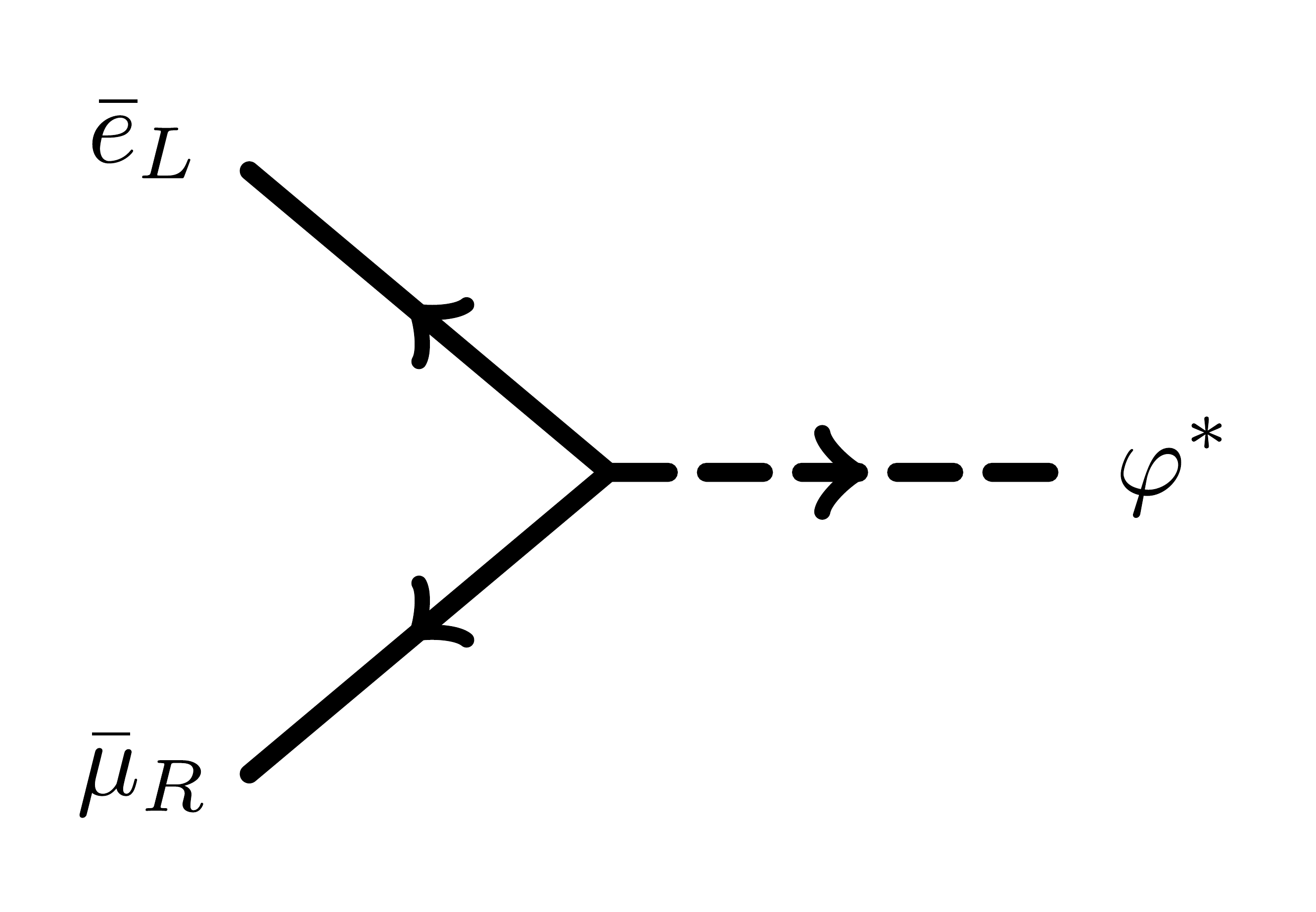}
  \;
  \includegraphics[width=.22\textwidth]{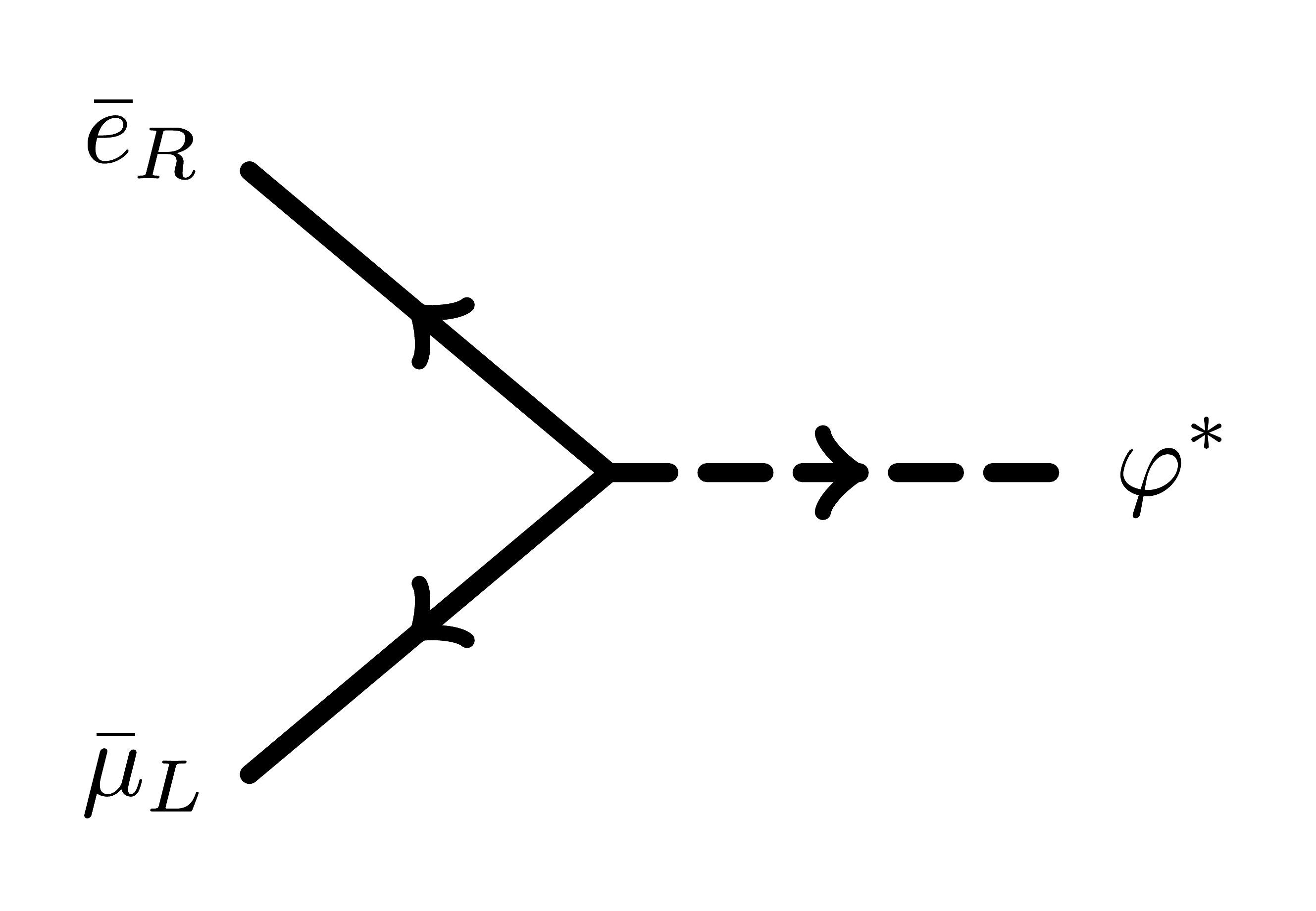}
  \caption{Vertices from \eqref{eq:Weyl:verts} to those in \eqref{eq:Weyl:verts:star} where arrows on fermions represent helicity and the arrow on the scalar represents $L_e - L_\mu$ charge.}
  \label{fig:vertices}
\end{figure}

An example where this formalism is useful is to examine possible loop-level contributions to flavor-changing dipole operators. We can see that such operators necessarily connect fermions of the same chirality:
\begin{align}
  \bar e \sigma^{\mu \nu} \mu F_{\mu \nu} &= 
  \vcenter{
    \hbox{\includegraphics[width=.17\textwidth]{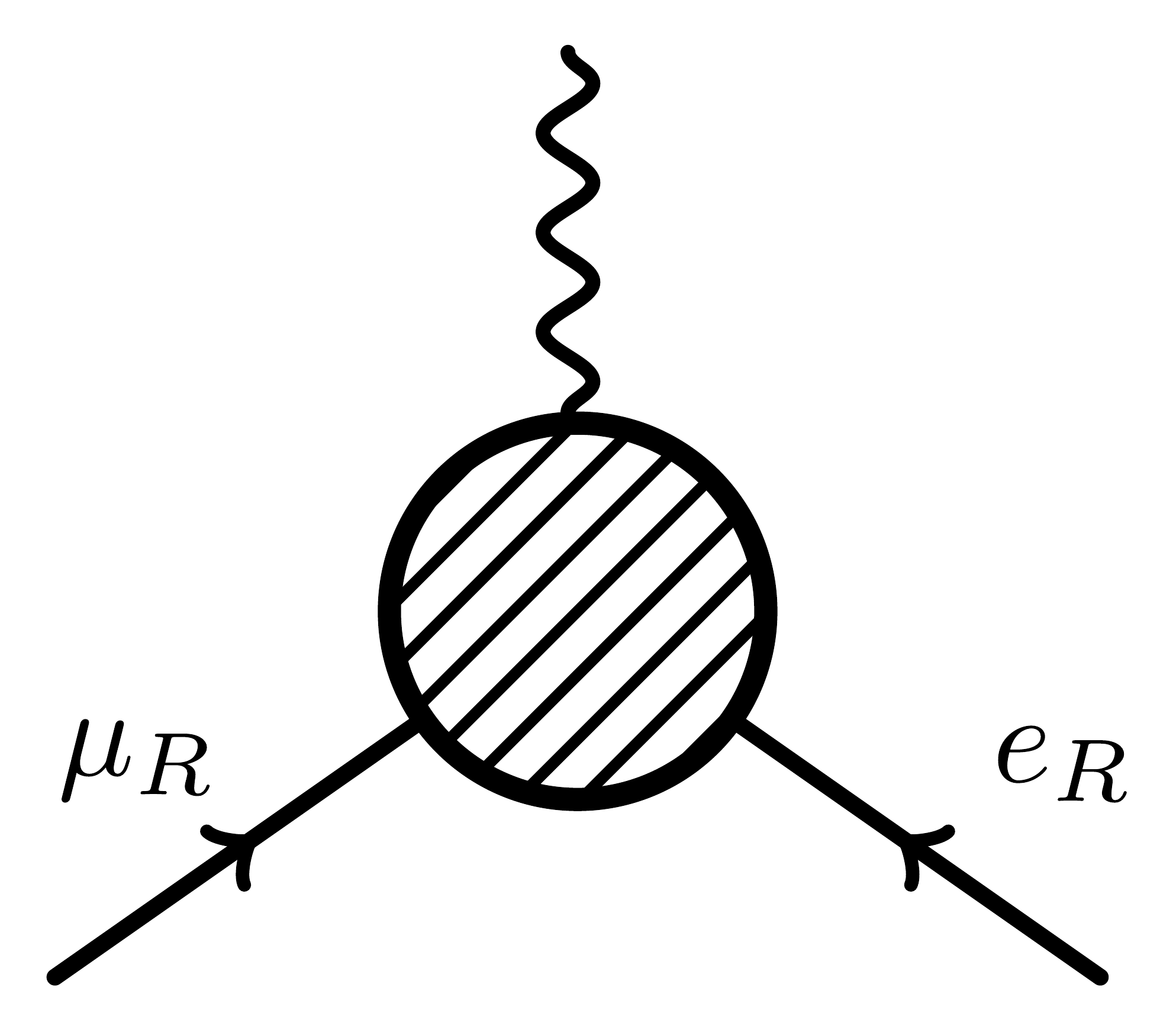}}
    }
    +
  \vcenter{
    \hbox{\includegraphics[width=.17\textwidth]{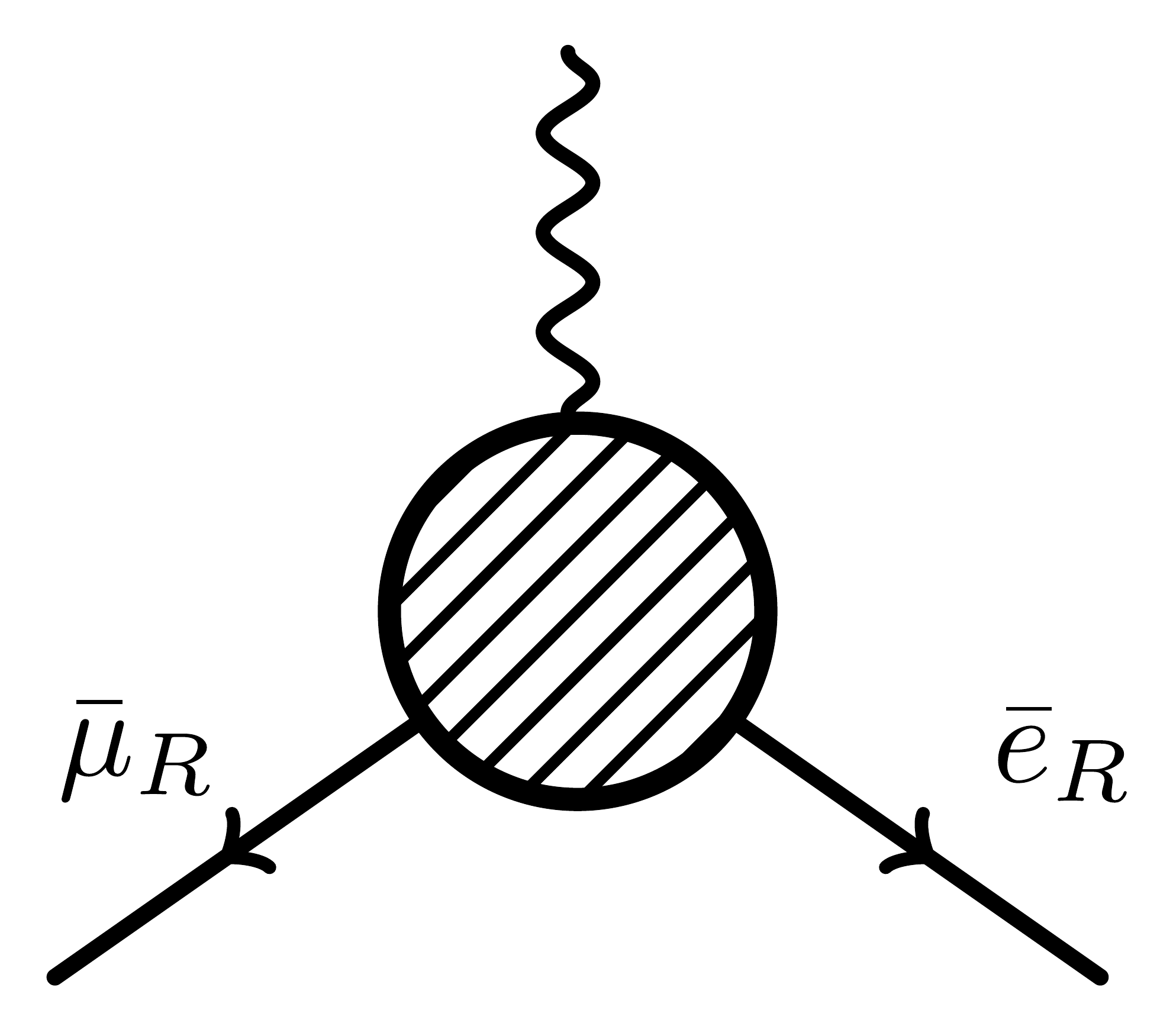}}
    } \ .
\end{align}
Because the lepton-flavor violating couplings of the $\varphi$ in (\ref{eq:Weyl:verts}--\ref{eq:Weyl:verts:star}), i.e.~Fig.~\ref{fig:vertices}, dipole operators in the limit of a single off-diagonal flavor coupling can only connect states of opposite chirality and different flavor. There is no one-loop diagram with an internal $\varphi$ for this process because of the spurious $L_e - L_\mu$ symmetry. This symmetry is violated by the $W$ interactions so that the leading diagram must contain additionally a loop with a neutrino--$W$ loop, as in the Standard Model process.
One finds that the leading contribution to $\mu\to e\gamma$ from the $\varphi$ is suppressed by an additional loop factor compared to the already tiny Standard Model term.

\bibliographystyle{utphys}
\bibliography{LFVbib}

\end{document}